\documentclass[10pt]{iopart}

\usepackage{iopams}
\usepackage[english]{babel}
\usepackage[utf8]{inputenc}
\usepackage{graphicx}
\usepackage{fancyhdr}
\usepackage[active]{srcltx}
\usepackage{setspace}
\usepackage{float}
\usepackage{color}
\usepackage{multirow}
\usepackage[hidelinks]{hyperref}
\usepackage{xcolor}

\hypersetup{
    colorlinks,
    linkcolor={red!50!black},
    citecolor={blue!50!black},
    urlcolor={blue!80!black}
}

\newcommand{\PEFJ}{{\normalfont P E F\lowercase{aria} J\lowercase{unior}} \lowercase{et al.}}

\begin{document}

\title[\PEFJ]{Proximity-enhanced valley Zeeman splitting at the WS$_2$/graphene interface}

\author{Paulo E. Faria~Junior}
\address{Institute for Theoretical Physics, University of Regensburg, 93040 Regensburg, Germany}
\ead{paulo-eduardo.faria-junior@ur.de; fariajunior.pe@gmail.com}

\author{Thomas Naimer}
\address{Institute for Theoretical Physics, University of Regensburg, 93040 Regensburg, Germany}

\author{Kathleen M. McCreary}
\address{Materials Science and Technology
Division, Naval Research Laboratory, Washington, Washington DC 20375, USA}

\author{Berend T. Jonker}
\address{Materials Science and Technology
Division, Naval Research Laboratory, Washington, Washington DC 20375, USA}

\author{Jonathan J. Finley}
\address{Walter Schottky Institut and TUM School of Natural Science, Technische Universität München, Am Coulombwall 4, 85748 Garching, Germany}

\author{Scott A. Crooker}
\address{National High Magnetic Field Laboratory, Los Alamos, New Mexico 87545, USA}

\author{Jaroslav Fabian}
\address{Institute for Theoretical Physics, University of Regensburg, 93040 Regensburg, Germany}
\ead{jaroslav.fabian@ur.de}

\author{Andreas V. Stier}
\address{National High Magnetic Field Laboratory, Los Alamos, New Mexico 87545, USA}
\address{Walter Schottky Institut and TUM School of Natural Science, Technische Universität München, Am Coulombwall 4, 85748 Garching, Germany}
\ead{andreas.stier@wsi.tum.de}

\maketitle


\begin{abstract}

The valley Zeeman physics of excitons in monolayer transition metal dichalcogenides provides valuable insight into the spin and orbital degrees of freedom inherent to these materials. Being atomically-thin materials, these degrees of freedom can be influenced by the presence of adjacent layers, due to proximity interactions that arise from wave function overlap across the 2D interface. Here, we report 60 T magnetoreflection spectroscopy of the A- and B- excitons in monolayer WS$_2$, systematically encapsulated in monolayer graphene. While the observed variations of the valley Zeeman effect for the A- exciton are qualitatively in accord with expectations from the bandgap reduction and modification of the exciton binding energy due to the graphene-induced dielectric screening, the valley Zeeman effect for the B- exciton behaves markedly different. We investigate prototypical WS$_2$/graphene stacks employing first-principles calculations and find that the lower conduction band of WS$_2$ at the $K/K'$ valleys (the $CB^-$ band) is strongly influenced by the graphene layer on the orbital level. This leads to variations in the valley Zeeman physics of the B- exciton, consistent with the experimental observations. Our detailed microscopic analysis reveals that the conduction band at the $Q$ point of WS$_2$ mediates the coupling between $CB^-$ and graphene due to resonant energy conditions and strong coupling to the Dirac cone. Our results therefore expand the consequences of proximity effects in multilayer semiconductor stacks, showing that wave function hybridization can be a multi-step process with different bands mediating the interlayer interactions. Such effects can be exploited to resonantly engineer the spin-valley degrees of freedom in van der Waals and moiré heterostructures.

\end{abstract}

{\bf Keywords:} TMDs, graphene, valley Zeeman effect, proximity

\maketitle

\ioptwocol


\section*{I. Introduction}

Van der Waals layered materials allow for the assembly of intentionally designed stacks with a dedicated topology or functionality\cite{Geim2013, Song2018, Sierra2021, Slobodkin2020PRL, Plankl2021NatPhot, Barre2022Science}. A key concept in this regard are proximity effects, where properties of a material or an ordered state are transferred from one layer to another without strongly affecting its electronic structure\cite{Zutic2019}. Nevertheless, a direct overlap of the wave functions in the adjacent layers is required for the proximity effect to take place. For example, pristine graphene (Gr) is a gapless Dirac semimetal with a linear dispersion relation around the $K/K'$ points and negligible spin-orbit coupling (SOC)\cite{Castro2009} while proximity effects from adjacent materials may significantly tailor its properties to acquire a positive or negative mass\cite{Avsar2020}, spin polarization and spin-orbit coupling\cite{Avsar2014, Gmitra2015PRB, Gmitra2016PRB, Avsar2017ACSNano, Luo2017NL, Cummings2017PRL, Fulop2021npj2D} or superconductivity\cite{Ojeda2009,Natterer2016,Burset2008,Hogl2020}. 

In recent years, the interface of Gr with a monolayer transition-metal dichalcogenide (ML TMD) has received wide attention. This system provides the appealing situation of creating a heterojunction between a material with strong SOC (the ML TMD) and long spin lifetimes (Gr). With respect to spin physics, particular interest was given to the topic of proximity induced spin-orbit coupling in Gr due to the strong SOC of the TMD\cite{Gmitra2015PRB, Gmitra2016PRB, Avsar2017ACSNano, Luo2017NL, Cummings2017PRL, Frank2018PRL, Li2019PRB, David2019PRB, Avsar2020, Fulop2021npj2D, Naimer2021PRB, Peterfalvi2022PRR, Pezo2021TDM}. Typically, proximity effects in the Gr community are theoretically modeled as a perturbation to the low energy model Hamiltonian of the Dirac cone\cite{Sierra2021}, while keeping the wave functions unmodified (taken as purely p$_\textrm{z}$ orbitals in Gr).

From the perspective of the optical properties of ML TMDs, engineering the dielectric environment with hexagonal boron nitride\cite{Stier.2016c,Wierzbowski2017,Cadiz.2017, Raja2019, Waldecker2019PRL, Goryca.2019} or Gr\cite{Raja2017, Lorchat2020NatNano, Waldecker2019PRL}, has been shown to be a very efficient path for tunable modification of the exciton binding energy ($E_B$) or sub nm lateral modulation of the TMD band gap ($E_g$)\cite{Raja2017,Tebbe2022}. This unprecedented degree of freedom provides novel functionality with respect to lateral heterojunctions, a technological feature which is very hard to realize in conventional semiconductor technology. In terms of conventional semiconductor spin physics, changes in the band gap are associated with strong modifications of the band $g$-factors, particularly in materials with strong SOC\cite{Enderlein1997, Yu2001, Winkler2003}. Therefore, it is expected that the ability to modify the band gap in the TMDs provides a path to tailor its exciton valley Zeeman effect. Recent state-of-the-art first principles calculations\cite{Wozniak2020PRB, Deilmann2020PRL, Forste2020NatComm, Xuan2020PRR} have shown how to properly evaluate the orbital angular momentum contribution to the exciton valley Zeeman effect in ML TMDs, taking into account the Bloch functions of conduction and valence band electrons. Interestingly, the exciton $g$-factor depends only weakly on the band gap, while the band $g$-factors are indeed more sensitive to $E_g$\cite{Wozniak2020PRB}, as expected from conventional III-V semiconductors within the k.p framework\cite{Roth1959PR, Winkler2003}. While the valley Zeeman physics in intrinsic monolayers\cite{Wozniak2020PRB, Deilmann2020PRL, Forste2020NatComm, Xuan2020PRR, Covre2022Nanoscale, Blundo2022PRL, FariaJunior2022NJP} and hetero/homo-bilayers\cite{Wozniak2020PRB, Xuan2020PRR, Gillen2021pssb, Forg2021NatComm, Heissenbuettel2021NL, Zhao2022, Kipczak2022} is relatively well understood based on the recent ab initio developments, the influence of finite carrier density\cite{Arora2016, Roch2019, Roch2020, Klein2021, Klein2022, Grzeszczyk2021, Li2021NatNano} or the evolution of the (in-plane) spin and orbital degrees of freedom in multilayered van der Waals heterostructures\cite{Raiber2022NatComm} still require further work.

In this study, we systematically investigate the dependence of the 1s exciton valley Zeeman $g$-factor on van der Waals heterostructures of monolayer WS$_2$ with graphene. We performed circularly polarized magneto-reflection spectroscopy up to 60 T on large area films of ML WS$_2$ grown by chemical vapor deposition (CVD). The films were transferred either directly on SiO$_2$, or were single/double encapsulated with monolayer Gr. Clear valley splittings for the A- and B- excitons ($X^{A,B}$) are observed, providing measurements of the associated exciton $g$-factors for each assembled structure. While the $g$-factor of the $X^A$ varies smoothly and consistently with the band gap renormalization and size of the exciton wave function, surprisingly the $g$-factor of $X^B$ varies more strongly in magnitude and distinctly non-monotonic. We explore the microscopic origin of this behavior with detailed first-principles calculations on several prototypical WS$_2$/Gr heterostructures with different stacking, shifts and twist angles. We show that the WS$_2$ conduction bands at the $Q$ point mediate the interaction between the lower conduction band ($CB^-$) at the $K$ point and the graphene Dirac cone, leading to distinct changes on the orbital degree of freedom of $CB^-$ and, consequently, on the B- exciton $g$-factor. This mediated coupling happens because of the energetic alignment of the conduction band $Q$ and $CB^-$ at $K$, which is absent in Mo-based TMDs. Furthermore, our quantitative account of the dielectric screening effects of Gr to the exciton $g$-factors (reduction of the band gap and localization of exciton wave function) strengthens our picture that the non-monotonic variations in $g_A - g_B$ are indeed signatures of the complex interlayer hybridization between WS$_2$ and Gr. Our results therefore expand the concept of proximity effects, revealing that interlayer wave function hybridization of adjacent crystalline layers can happen at different levels (mediated by different energy bands), a concept that is crucial for understanding the spin-valley physics of van der Waals and moiré heterostructures.


\section*{II. Experimentally determined valley Zeeman effect in WS$_2$/graphene systems}

\begin{figure}[htb]
\begin{center} 
\includegraphics[width=0.49\textwidth]{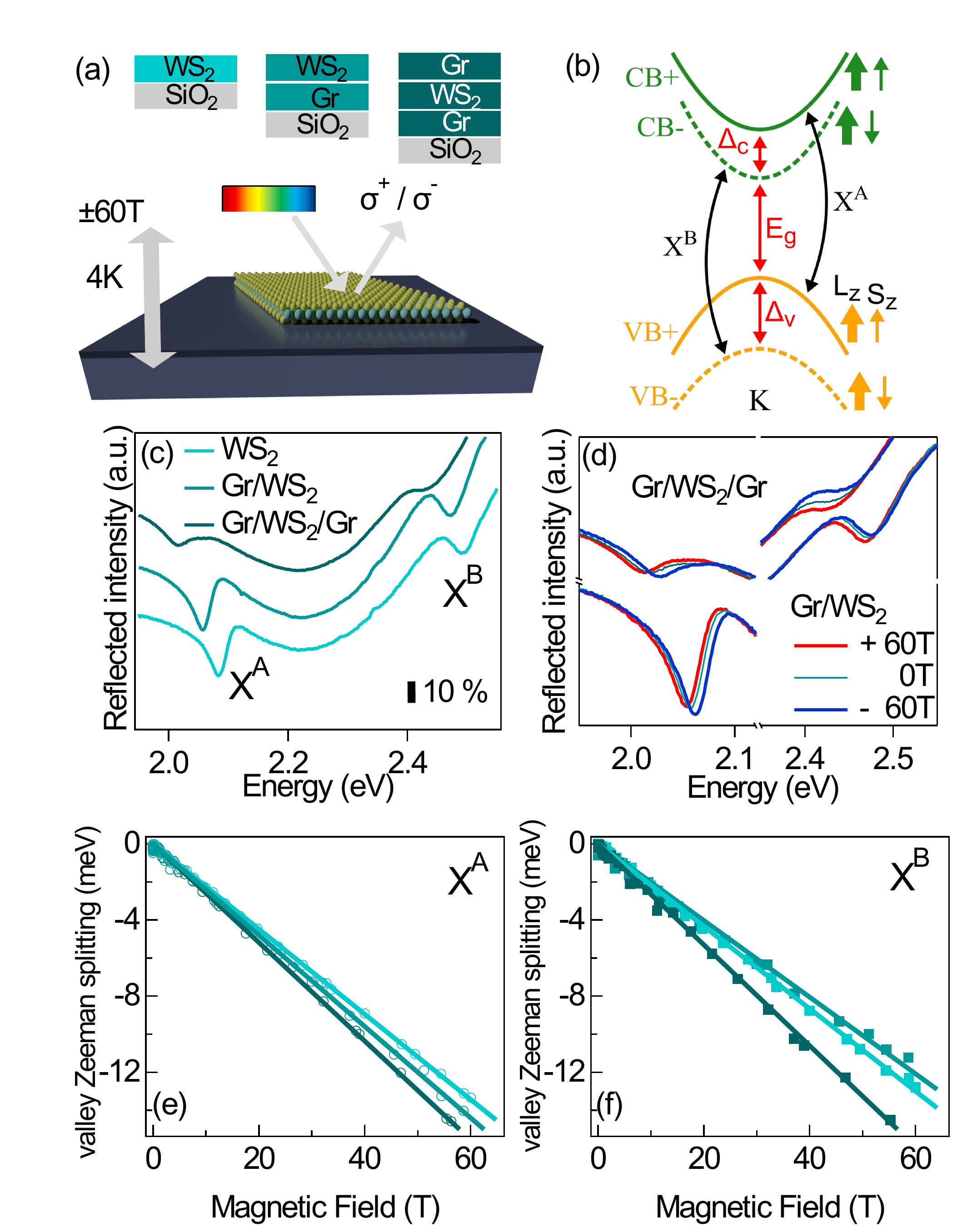}
\caption{(a) Schematic of the low temperature magneto-reflectivity experiment and investigated sample stacks. Unpolarized light is focused on the sample with a single aspheric lens and reflected from the sample surface into a collection multimode fiber. Left- and right circularly polarized light is analyzed through a thin film polarizer. (b) Diagram of the conduction and valence bands close to the $K$ point of the WS$_2$ Brillouin zone, showing A and B exciton transitions ($X^{A,B}$) and the associated orbital angular momentum ($L_Z$) and spin ($S_Z$) configurations. Spin up/down bands are separated by spin-orbit splitting $\Delta_{c,v}$.  (c) Zero $B$-field reflection spectra at $T=4K$ of the samples. Lines are offset for clarity. (d) $X^{A,B}$ resonances of single and double encapsulated monolayer WS$_2$ at $B=\pm 60 \; T$. (e) Valley Zeeman splitting of $X^{A}$, and (f) $X^{B}$. While the valley Zeeman splitting of $X^{A}$ evolves smoothly with increased dielectric screening, the $X^{B}$ valley Zeeman splitting depends non-monotonically on encapsulation.}
\label{fig_1}
\end{center}
\end{figure}

We depict a schematic of the experiment and the investigated sample stacks in Figure~\ref{fig_1}(a). Large-area monolayer films of WS$_2$ were grown by CVD on SiO$_2$/Si or graphene substrates\cite{McCreary2014AFM, McCreary2016SR2, Paradisanos.2020}. The monolayer nature and high quality of these samples were confirmed by photoluminescence and Raman spectroscopy maps\cite{McCreary2016SR, McCreary2016APL, Stier.2016}. The as-grown monolayers were transferred from the growth substrate to a Si/SiO$_2$ substrate via standard wet transfer methods. For the doubly graphene encapsulated WS$_2$ layer, a top graphene layer was wet-transferred on the WS$_2$/graphene CVD film on the target substrate. Magneto-reflectance studies were performed at cryogenic temperatures ($T = 4 \; \textrm{K}$) in a capacitor-driven 65 T pulsed magnet at the National High Magnetic Field Laboratory in Los Alamos. Details about the experimental setup can be found in the Supplemental Note I\cite{SM} and in Ref.\cite{Stier.2016c}. A total of three samples of each batch were investigated with two separate spots on each of the three samples. Data shown in this manuscript are typical for each batch and the reported error bars derive from averages of all experiments done on each batch. 

We investigate the effect of graphene encapsulation on the exciton binding energy $E_B$ and in particular the evolution of the valley Zeeman effect\cite{Xiao.2012} of the A and B excitons, depicted in the single-particle energy diagram of the conduction and valence band in monolayer TMDs close to the $K$ point of the Brillouin zone [Fig.~\ref{fig_1}(b)]. These excitons are observed as dips in the smooth reflection spectra depicted in Fig.~\ref{fig_1}(c) close to 2.05 eV ($X^A$) and 2.45 eV ($X^B$), respectively. No charged exciton features can be observed, confirming that the samples are close to intrinsic\cite{Chernikov.2015}, although no active carrier control through gates has been employed here. The energy difference between $X^A$ and $X^B$ at zero magnetic field is given by the difference in the respective exciton binding energy and the SOC-induced splitting of the conduction ($\Delta_c$) and valence band ($\Delta_v$). As mentioned in the introduction, graphene encapsulation successively introduces screening for the interband transitions in the TMD ML. This directly affects the band gap and $E_B$, causing the excitonic transitions to shift with varying encapsulation [see Fig.~\ref{fig_1}(c)]. The effect of the increased screening on the 2D excitons can be probed in high field magnetospectroscopy.

In a magnetic field, the exciton energies shift following the relation 
\begin{equation}
\Delta E_{A,B}(B)=\sigma_{A,B} B^2+ \frac{1}{2} \tau g_{A,B}\mu_B B,
\end{equation}
where the diamagnetic shift, 
\begin{equation}
E_{dia}=\sigma B^2=e^2\langle r^2 \rangle_{1s}B^2/8 m_r,
\end{equation}
the reduced mass of the exciton is $m_r=(1/m_e+1/m_h)^{-1}$ and the exciton ground state rms size is $(r_1=\sqrt{\langle r^2 \rangle_{1s}})$. 
$\tau=\pm1$ is the $K/K'$ valley index, $\mu_B = e\hbar/2m_0$ is the Bohr magneton,  and $g_{A,B}$ are the valley Zeeman $g$-factors of the excitons, related to the relevant energy bands by
\begin{eqnarray}
g_{A} & =& 2\left[g_{z}(\textrm{CB}^+,\textrm{K})-g_{z}(\textrm{VB}^+,\textrm{K})\right]\nonumber \\
g_{B} & = &2\left[g_{z}(\textrm{CB}^-,\textrm{K})-g_{z}(\textrm{VB}^-,\textrm{K})\right]
\label{eq:gAB}
\end{eqnarray}
with $g_{z}(n,\vec{k})$ being the out-of-plane g-factor of the Bloch band $n$ with wave vector $\vec{k}$ [see the relevant bands in Fig.~\ref{fig_1}(b)].

\begin{table}[htb]
\caption{Experimentally determined g-factors from the linear fittings of Figs.~\ref{fig_1}(e-f). The experimental error is $0.1$.}
\begin{center}
\begin{tabular}{cccc}
\hline 
\hline
 & WS$_2$ & Gr/WS$_2$ & Gr/WS$_2$/Gr\tabularnewline
\hline 
$g_{A}$ & -3.8 & -4.0 & -4.2\tabularnewline
$g_{B}$ & -3.7 & -3.3 & -4.5\tabularnewline
$g_{A}-g_{B}$ & -0.1 & -0.7 & 0.3\tabularnewline
\hline
\hline
\end{tabular}
\end{center}
\label{tab:g_exp}
\end{table}

Figure Fig.~\ref{fig_1}(d) shows the reflection spectra of $X^{A,B}$ for the single and doubly Gr-encapsulated WS$_2$ at zero and the maximum $\pm 60 \; \textrm{T}$ applied magnetic field. A valley Zeeman splitting of $\approx12$ meV for each, $X^{A,B}$ resonance, analyzed in detail below, is observed. Both resonances shift to lower (higher) energy in positive (negative) fields, indicating a negative $g$-factor. The spectral features were fit using complex (absorptive and dispersive) Lorentzian lineshapes to extract the transition energy. Although the exact position of each resonance at zero magnetic field is difficult to determine precisely due to the smoothly varying background, the magnetic field dependent shifts can be exactly determined, as the background is unaffected during the magnetic field pulse. 
As such, the diamagnetic shift and the valley Zeeman splitting can be simply determined from the average and difference of the exciton $B$-dependent shift in positive and negative fields. In the Supplemental Note II - IV\cite{SM}, we provide further details of the dielectric effects resulting in the decrease of the exciton binding energy.
Figures~\ref{fig_1}(e,f) show the deduced valley Zeeman splittings for $X^A$ and $X^B$, respectively. While the magnetic moment of the A-exciton smoothly increases in magnitude with increasing encapsulation, surprisingly, $g_B$ evolves in a distinctly non-monotonic manner. This can be seen in the raw data of Fig.~\ref{fig_1}(f), where clearly the B-exciton of the double-encapsulated WS$_2$ splits significantly more than that of the single sided encapsulated TMD. The collected $g$-factor values are given in Table~\ref{tab:g_exp}. The unexpected behavior of the markedly different evolution of the exciton $g$-factors, best highlighted through the difference $g_A - g_B$, is the key experimental result of this study and can only be explained by taking interface hybridization of the wave functions into account.


\section*{III. Proximitized valley Zeeman physics in WS$_2$/graphene heterostructures}

\begin{figure*}[htb]
\begin{center} 
\includegraphics{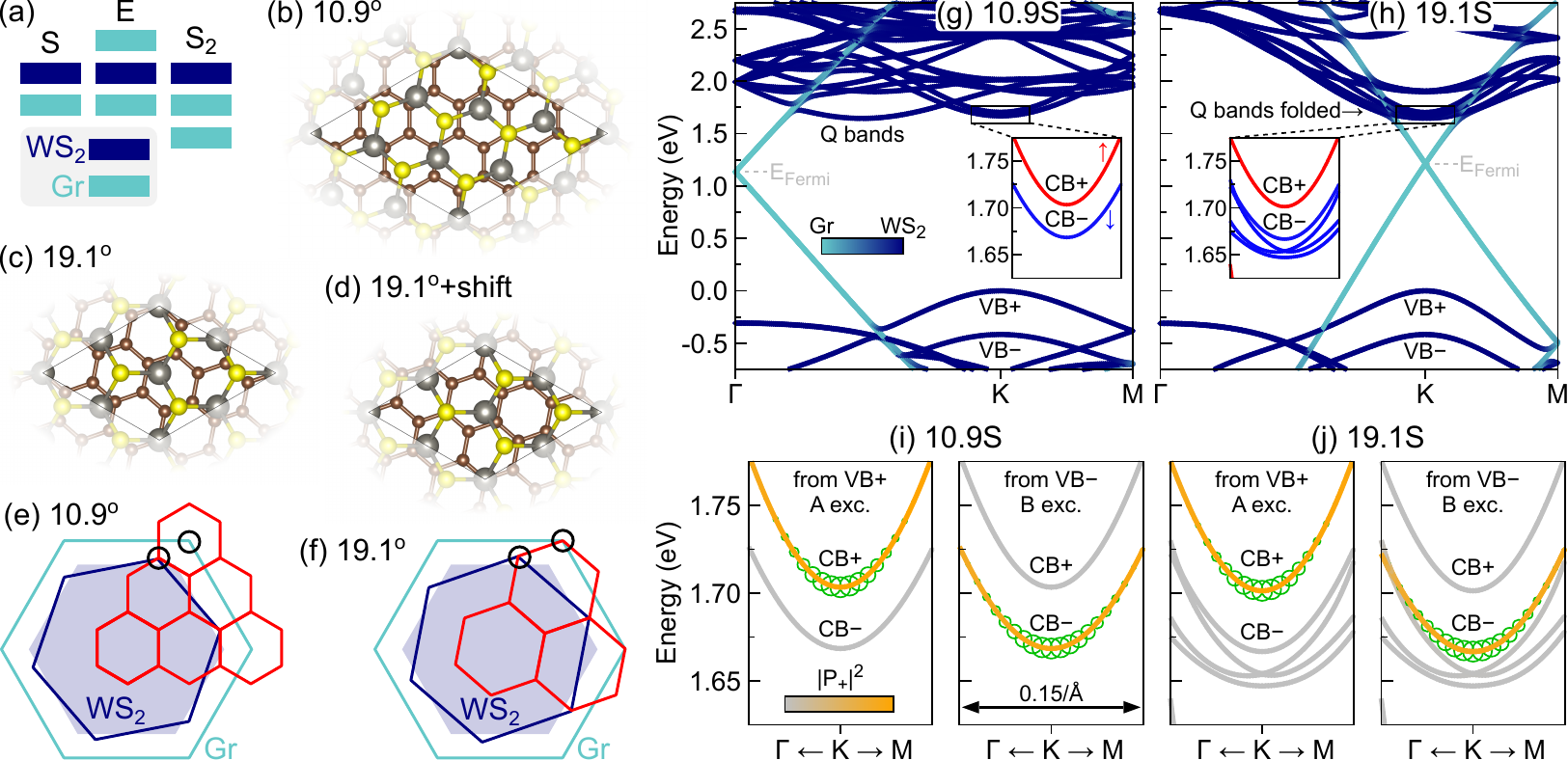}
\caption{(a) Different configurations of the WS$_2$/Gr van der Waals heterostructures considered in our study: S (substrate), E (encapsulated) and S$_\textrm{2}$ (bilayer substrate). Commensurate supercells for the (b) 10.9$^\circ$, (c) 19.1$^\circ$~and (d) 19.1$^\circ$~with a shift in the S configuration. Hexagonal Brillouin zones for the (e) 10.9$^\circ$ and (f) 19.1$^\circ$~twist angles (graphene has the largest Brillouin zone, followed by WS$_2$ and by the supercell). The circles indicate the folding of the TMD and Gr K valleys to the supercell Brillouin zone. Electronic band structures with Gr/WS$_2$ layer decomposition for the (g) 10.9S and (h) 19.1S studied cases. The Fermi energy is shown by the horizontal dashed line. The relative positions of the TMD and Gr $K$ valleys are consistent with figures (e-f). The insets show a zoom of the conduction bands $CB^\pm$ with their associated spin expectation value (red $\uparrow$ and blue $\downarrow$, calculated via Eq.~\ref{eq:Sz}). In-plane dipole transition amplitudes around the $K$ valley for optical transitions originating from the WS$_2$ top valence bands, $VB^\pm$, for the (i) 10.9S and (j) 19.1S cases. The dipole transition amplitude is given by $\left| \textrm{P}_+ \right|^2=\left|\left\langle v,\vec{k}\left|\hat{\sigma}_{+}\cdot\vec{p}\,\right|c,\vec{k}\right\rangle \right|^{2}$, with $\hat{\sigma}_{+}=\left(\hat{x}+i\hat{y}\right)/\sqrt{2}$, and is normalized by the $VB^+$ $\leftrightarrow$ $CB^+$ value directly at the $K$ point. The (green) open circles depict the probability density of the exciton envelope function (details in Supplemental Note IV\cite{SM}).}
\label{fig_2}
\end{center}
\end{figure*}

The interlayer hybridization of different layered materials is typically dependent on several parameters such as the lattice mismatch, the twist angle, the atomic registry, and so on. To investigate proximity effects in the valley Zeeman physics of WS$_2$/Gr van der Waals heterostructures, we consider several prototypical systems with different stackings and twist angles, calculated from first principles. The electronic properties are calculated via density functional theory (DFT) using the an all-electron full-potential implementation within WIEN2k\cite{wien2k}, one of the most accurate DFT codes available\cite{Lejaeghere2016Science}, which has been successfully applied to investigate the microscopic nuances of SOC and spin-phenomena in 2D materials and their van der Waals heterostructures (including, but not limited to, Gr and TMDs)\cite{Gmitra2009PRB, Konschuh2010PRB, Gmitra2015PRB, Gmitra2016PRB, Kurpas2016PRB, Kurpas2019PRB, FariaJunior2019PRB, FariaJunior2022NJP}. The computational details can be found in the Supplemental Note V\cite{SM}. We note that previous DFT works on TMD and Gr systems have considered different commensurate structures, with different strain values and twist angles\cite{Ratha2014APL, Gmitra2016PRB, Gmitra2015PRB, Paradisanos2020APL, Naimer2021PRB, Pezo2021TDM, Yang2022Photonics}, and therefore there is no unique recipe on how to construct the van der Waals heterostructures of TMDs and Gr. An important point for our analysis is that only Gr is strained, so that the observed changes in the WS$_2$ $g$-factors arise solely from the interlayer coupling between the two materials. Strain effects in the $g$-factors of monolayer TMDs have already been investigated in Ref.~\cite{FariaJunior2022NJP}, while strain effects in Gr are known to influence the Fermi velocity of the Dirac cone\cite{Choi2010PRB}, leaving other features practically unaltered.

In Fig.~\ref{fig_2}(a), we depict the TMD/Gr systems considered here and contemplate three different cases: graphene as a substrate (S), graphene encapsulation (E) and bilayer graphene as a substrate (S$_\textrm{2}$). The S and E cases are chosen to mimic the Gr/WS$_2$ and the Gr/WS$_2$/Gr experimental samples discussed in Section II. The S2 configuration has no experimental counterpart in this study, but it is a typical structure considered by the Gr community as a platform for proximity-induced SOC effects\cite{Gmitra2017PRL, Lee2020PRL, Amann2022PRB, Zhang2023Nature}. In  Fig.~\ref{fig_2}(b-d), we show the atomic structures of the WS$_2$/Gr supercells for the different twist angles considered in this study, i.e., 10.9$^\circ$~in Fig.~\ref{fig_2}(b), 19.1$^\circ$~in Fig.~\ref{fig_2}(c) and 19.1$^\circ$~with an in-plane shift in Fig.~\ref{fig_2}(d).  The encapsulated systems have an additional graphene layer on top of the TMD while the bilayer graphene cases have a second graphene layer below, either with AB (Bernal) or AA stacking. Particularly, these two twist angles we consider here (10.9$^\circ$ and 19.1$^\circ$) provide a relatively small number of atoms (see Supplemental Note V\cite{SM}) and different folding of $k$ points (see next paragraph).

In the reciprocal space, the twist angle defines how different $k$ points of the individual layers are mapped, or folded, to the reciprocal space of the supercell. The first Brillouin zone (BZ) of Gr, WS$_2$ and the resulting supercell are shown in Figs.~\ref{fig_2}(e,f) for 10.9$^\circ$~and 19.1$^\circ$, respectively. Importantly, the relative alignment of the $K$ points in TMD and Gr happen at different positions for the two different angles, thus serving as the limiting cases for our analysis of the interlayer coupling effect. In realistic systems with long moiré lengths in real space (10s of nm), many more BZs of the supercell would fit in the TMD and Gr BZs and therefore many more $k$ points of the individual layers would be folded to a single $k$ point of the supercell BZ.

The folding of the different $k$ points, as well as the interactions between the layers, can be directly seen in the electronic properties (akin to electronic levels interacting via weak periodic potentials\cite{Ashcroft1976}). In Figs.~\ref{fig_2}(g,h) we show the layer-resolved band structures for 10.9S and 19.1S cases, respectively. It is evident that these two choices of twist angle provide different alignments between the low energy TMD bands and the Dirac cone, i.e., for 10.9$^\circ$ (19.1$^\circ$) the TMD bands are located at the $K$ point and the Dirac cone is located at the $\Gamma$ ($K$) point, in agreement with the expectations from Figs.~\ref{fig_2}(e,f). Furthermore, besides the folding of $k$ points, the interlayer coupling between WS$_2$ and Gr induces splittings to the energy bands, more visible in the regions where dark and bright blue regions overlap. The valence bands $VB^\pm$ are nicely isolated from the other bands, whereas the conduction bands $CB^\pm$ (indicated by the black rectangles, expanded as the inset) show a different behavior, particularly, the folding of the $Q$ point conduction bands to the $K$ point in the 19.1S case. We highlight the spin orientation of these folded $Q$ bands in the inset and they exhibit the same spin direction as in the $CB^-$.

\begin{figure*}[htb]
\begin{center} 
\includegraphics{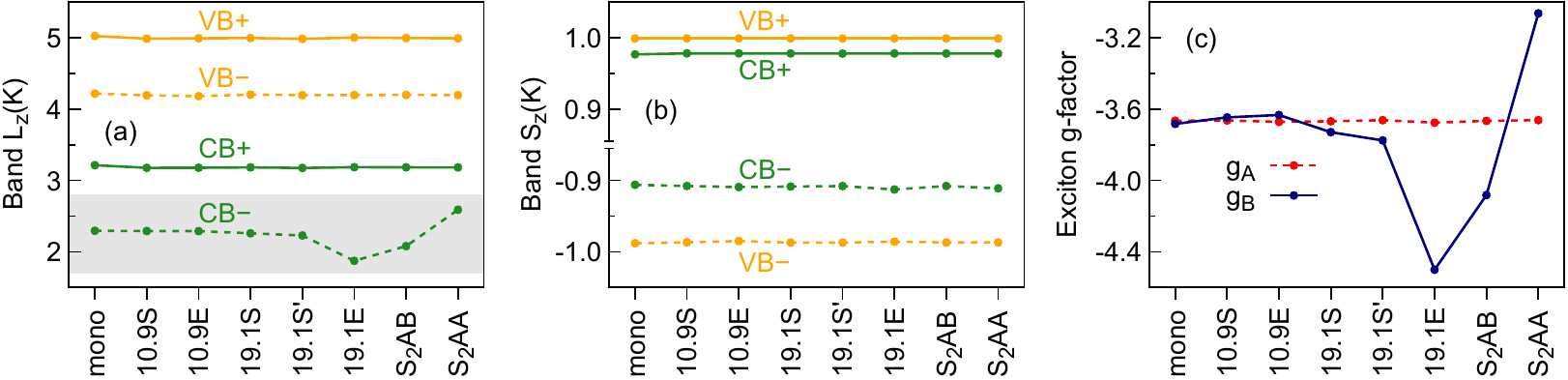}
\caption{(a) Orbital angular momentum, $L_z$, and (b) spin angular momentum, $S_z$, for the relevant energy bands $CB^\pm$ and $VB^\pm$. The shaded area in panel (a) highlight the origin of the proximity valley Zeeman, an effect on the orbital level that modifies the orbital angular momentum of the lowest conduction band $CB^-$. The line color and style in panels (a-b) follow the schematics of Fig.~\ref{fig_1}(b). (c) Calculated exciton g-factors, $g_A$ and $g_B$, for the different WS$_2$/Gr systems considered. Because of the changes in $L_z$ of $CB^-$, only the $g_B$ is modified, leaving $g_A$ essentially unaltered. For the x-axis, the 19.1S' case corresponds to 19.1$^\circ$~with an in-plane shift [Fig.~\ref{fig_2}(d)] and the S$_2$AB or S$_2$AA cases correspond to the 19.1S case with a second Gr layer underneath in order to generate a bilayer graphene substrate with AB (Bernal) or AA stacking.}
\label{fig_3}
\end{center}
\end{figure*}

In Figs.~\ref{fig_2}(i,j) we focus on the energy scale of the conduction bands $CB^\pm$ that are contained in the insets in Figs.~\ref{fig_2}(g,h). The color code, from gray to orange, indicates the amplitude of the dipole transition from the $VB^\pm$ with $\sigma_+$ circularly polarized light, in order to identify the optical transitions that contribute to $X^{A,B}$. Our calculations show that the selection rules still hold in the heterostructure and, more importantly, that the $Q$ point folded bands are optically inactive. The exciton spreading in $k$-space is also shown (using the calculated values of the Gr/WS$_2$ case, discussed in Section II with details in Supplemental Note IV\cite{SM}) and reveals that only a small region around the $K$ point is relevant, in line with robust GW-BSE calculations in bare monolayers\cite{Katznelson2022MatHor}. Therefore, we can investigate the signatures of the WS$_2$/Gr hybridization by simply computing the $g$-factors of $CB^\pm$ and $VB^\pm$ directly at the $K$ points of the heterostructure.

In order to investigate the modified valley Zeeman physics due to the microscopic effects of WS$_2$/Gr interlayer hybridization and its signature on the experimentally observed exciton $g$-factors (given in Eq.~\ref{eq:gAB}), we must evaluate the Zeeman shift of the Bloch band (generally identified by the index $n$ and crystal momentum $\vec{k}$):
\begin{eqnarray}
E_{\textrm{zs}}(n,\vec{k}) & =\left[L_z(n,\vec{k})+S_z(n,\vec{k})\right]\mu_{B}B\nonumber \\
 & =g_z(n,\vec{k})\mu_{B}B \, ,
 \label{eq:EZ}
\end{eqnarray}
where $B$ is the external magnetic field (in the out-of-plane direction, $z$, as in Fig.~\ref{fig_1}(a), and $L_z(n,\vec{k})$, $S_z(n,\vec{k})$ and $g_z(n,\vec{k})$ are the orbital angular momenta, spin angular momenta and $g$-factor of the Bloch band, respectively. The calculation of the orbital angular momentum in the out-of-plane direction, $L_z(n,\vec{k})$ for a Bloch state is obtained via the summation-over-bands approach\cite{Roth1959PR, ChangNiu1996PRB, Wozniak2020PRB, Deilmann2020PRL, Forste2020NatComm, Xuan2020PRR, FariaJunior2022NJP}
\begin{equation}
L_z(n,\vec{k}) = \frac{1}{im_{0}} \sum_{m \neq n} \! \frac{P_{x}^{n,m,\vec{k}}P_{y}^{m,n,\vec{k}} \! - \! P_{y}^{n,m,\vec{k}}P_{x}^{m,n,\vec{k}}}{E(n,\vec{k})-E(m,\vec{k})} \, ,
\label{eq:Lz}
\end{equation}
in which $P_{\alpha}^{n,m,\vec{k}}=\left\langle n,\vec{k}\bigl| {p}_{\alpha}\bigr| m,\vec{k} \right\rangle$ ($\alpha=x,y,z$), with $\vec{p}$ being the momentum operator, and the spin angular momentum is calculated as
\begin{equation}
S_z(n,\vec{k}) = \left\langle n,\vec{k}\bigl|\hat{\sigma}_{z}\bigr|n,\vec{k}\right\rangle \, ,
\label{eq:Sz}
\end{equation}
with $\sigma_z$ the Pauli matrix acting on the spin-up and spin-down states of the spinorial Bloch state. We note that, because of time-reversal symmetry, the relation $O(n,-\vec{k}) = - O(n,\vec{k})$ holds for $O = L_z, S_z, g_z$. For further details on this theoretical approach applied to TMDs, we refer to Refs.\cite{Wozniak2020PRB, Deilmann2020PRL, Forste2020NatComm, Xuan2020PRR, FariaJunior2022NJP}. 

The calculated orbital and spin angular momenta of the $CB^\pm$ and $VB^\pm$ bands are shown in Figs.~\ref{fig_3}(a,b), respectively, for the investigated WS$_2$/Gr heterostructures. Our results reveal that $L_z$ of $CB^-$, highlighted by the gray area in Fig.~\ref{fig_3}(a), is distinctly modified, either increasing or decreasing with respect to the monolayer value depending on the particular system. The orbital angular momenta of the remaining bands ($VB^\pm$ and $CB^+$) barely changes. Additionally, we have not observed any changes in the spin angular momenta of the investigated bands. Combining the $L_z$ and $S_z$ of the energy bands, we can evaluate the A and B exciton g-factors via Eq.~\ref{eq:gAB}. The calculated values of $g_A$ and $g_B$ are shown in Fig.~\ref{fig_3}(c). While $g_A$ is essentially constant, $g_B$ is visibly changing (due to $L_z$ of $CB^-$), revealing that there are indeed sizable contributions to the valley Zeeman of excitons arising from the interlayer coupling between WS$_2$ and Gr layers. Therefore, we attribute the experimentally observed changes of $g_A - g_B$ (Table~\ref{tab:g_exp}) to manifestations of the interlayer hybridization at the Gr/WS$_2$ interface. We note that the monolayer thickness is slightly modified when we perform the atomic relaxation of the heterostructure, but this effect is not responsible for the drastic changes in $g_A - g_B$ (see Supplemental Note V\cite{SM} for this comparison). Furthermore, we emphasize that our numerical calculations for $L_z$ are fully converged with respect to the number of bands (see in Supplemental Note V\cite{SM} the comparison of monolayer and the 19.1E case).

\begin{figure}[htb]
\begin{center} 
\includegraphics{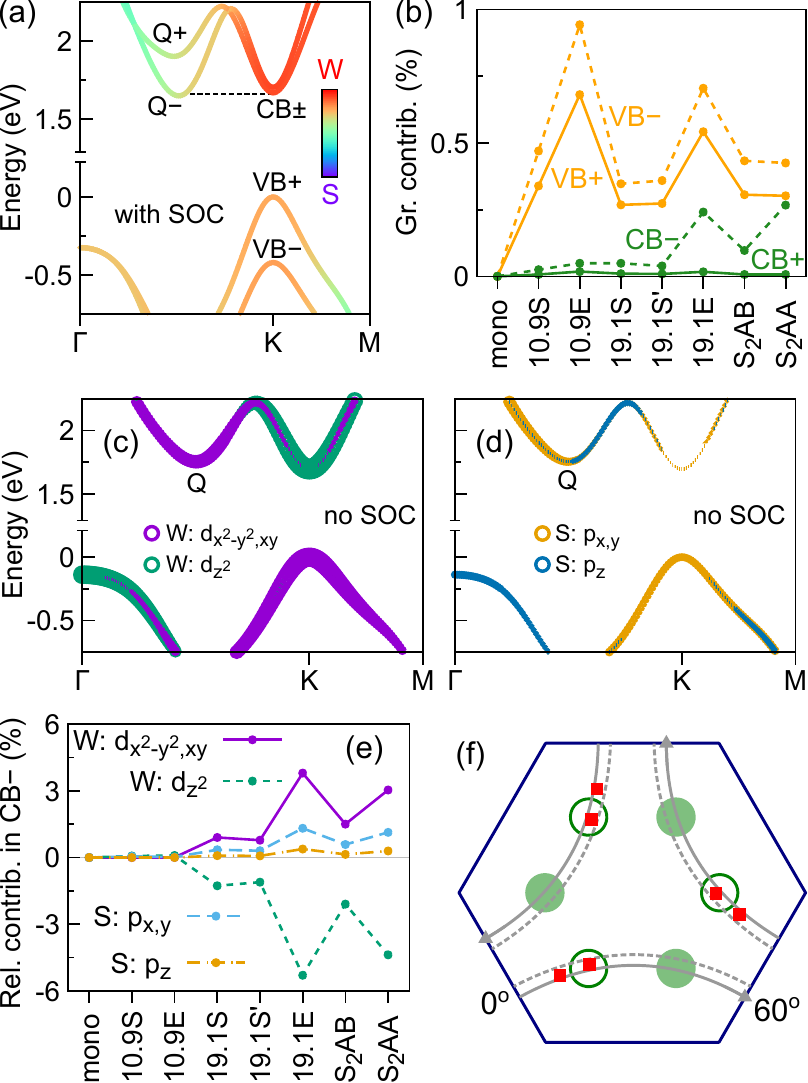}
\caption{(a) Band structure of WS$_2$ monolayer showing the spread of the wave function over W and S atomic spheres with SOC. The dashed horizontal line connecting the $Q^-$ and $CB^-$ bands indicate the resonant energy condition. Conduction bands at the $Q$ point have a much larger spin splitting (indicated by $Q^+$ and $Q^-$ bands). (b) Contribution of Gr atomic spheres (mainly p$_\textrm{z}$ orbitals) to the $CB^\pm$ and $VB^\pm$ bands of the studied WS$_2$/Gr systems. Band structure of WS$_2$ monolayer without SOC showing the majority orbital contributions from the (c) W atom (d$_{\textrm{x}^2-\textrm{y}^2\textrm{,xy}}$ and d$_{\textrm{z}^2}$) and (d) S atoms (p$_\textrm{x,y}$ and p$_\textrm{z}$). The size of circles indicate the contribution. (e) Orbital decomposition of $CB^-$ band for the WS$_2$/Gr heterostructures investigated. The detailed orbital analysis of the $CB^\pm$ and $VB^\pm$ is shown in Supplemental Note VI\cite{SM}. (f) First BZ of a TMD monolayer with the arrows indicating the position in $k$-space of the first-order coupling to the Dirac cone\cite{Koshino2015NJP, Li2019PRB, David2019PRB} as function of the twist angle (from 0$^\circ$~to 60$^\circ$). Open and closed circles indicate the Q$-$ bands with opposite spin orientation. The solid (dashed) lines indicate zero ($-5$\%) strain. The red squares indicate the 10.9$^\circ$ and 19.1$^\circ$. The 19.1$^\circ$ case lies within the range of the Q$-$ band.}
\label{fig_4}
\end{center}
\end{figure}

At a first glance, it may seem counter intuitive that the largest change of the $g$-factor originates from the orbital degree of freedom of the \textit{conduction} band $CB^-$. In Fig.~\ref{fig_4}(a) we show the wave function localization in the W and S atomic spheres for pristine WS$_2$ monolayer including SOC. The conduction bands, $CB^\pm$, at the $K$ point are highly localized at the W atoms, while valence bands, $VB^\pm$, are more delocalized across the layer towards the S atoms. Since $VB^\pm$ bands are more delocalized, one might expect them to be be more sensitive to the effect of the adjacent layers. In fact, the percentage of the wave function that ``leaks'' to the Gr layer is larger for $VB^\pm$ bands, as shown in Fig.~\ref{fig_4}(b). Nevertheless, the largest changes to the band $g$-factors are observed for $CB^-$, as shown in Fig.~\ref{fig_3}(a). Interestingly, the spin degree of freedom of $CB^-$ band in W-based TMDs is also the most affected by strain\cite{FariaJunior2022NJP}.

\begin{figure*}[htb]
\begin{center} 
\includegraphics{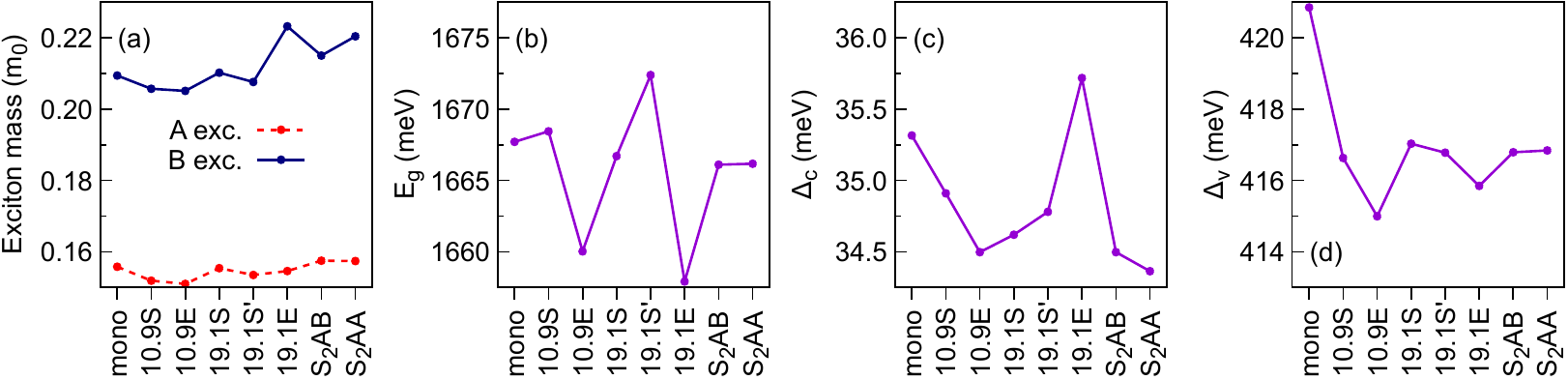}
\caption{(a) Effective masses for A and B excitons derived from first principles. (b) Band gap $E_g$, and spin-orbit splittings (c) $\Delta_c$ and (d) $\Delta_v$, as defined schematically in Fig.~\ref{fig_1}(b).}
\label{fig_5}
\end{center}
\end{figure*}

We now turn to the central aspect of our theoretical analysis, i. e., the microscopic mechanism behind the changes in $L_z$ of $CB^-$. For this purpose, we performed a detailed investigation of the orbital composition of the low energy bands $CB^\pm$ and $VB^\pm$. The monolayer WS$_2$ is summarized in Fig.~\ref{fig_4}(c,d) (SOC is neglected to simplify the visualization) and gives us the base values to compare with the investigated heterostructures. The conduction (valence) bands at the $K$ point are dominated by d$_{\textrm{z}^2}$ (d$_{\textrm{x}^2-\textrm{y}^2\textrm{,xy}}$) atomic orbitals of the W atom\cite{Xiao.2012} while conduction bands at the $Q$ point are mainly composed of d$_{\textrm{x}^2-\textrm{y}^2\textrm{,xy}}$ orbitals of the W atom. The p-like orbitals of the S atoms also provide a visible contribution, particular for the valence band at the $K$ point and conduction bands at the $Q$ point. In Fig.~\ref{fig_4}(e) we present the contribution of d-like (p-like) orbitals in W (S) atoms of the band $CB^-$ for the different WS$_2$/Gr systems, considering the monolayer case as reference. Surprisingly, we observe a decrease of the d$_{\textrm{z}^2}$ character of the W atom, accompanied with an increase of the d$_{\textrm{x}^2-\textrm{y}^2\textrm{,xy}}$ character of the W atom and p-like character (p$_\textrm{x,y}$ and p$_\textrm{z}$) of the S atoms, consistent with the orbital signature of the conduction band at the $Q$ point. The full orbital decomposition analysis is shown in the Supplemental Note VI\cite{SM}. For the 19.1$^\circ$ cases, besides the folding of the $Q$ bands to the $K$ point, these $Q$ bands also have a strong coupling to the Gr Dirac cone, as shown in Fig.~\ref{fig_4}(f) for the first-order Umklapp condition\cite{Koshino2015NJP, Li2019PRB, David2019PRB}, thus mediating the coupling between $CB^-$ and Gr states. We can summarize the hybridization mechanism as following: the TMD conduction bands at the $Q$ points (more delocalized than $K$ point bands) strongly couple with the Dirac cone in Gr (which has a large magnetic moment\cite{Frank2020PRB}) and then hybridize with the TMD conduction bands at the $K$ point. Because of the nearly resonant condition [horizontal dashed line in Fig.~\ref{fig_4}(a)] and spin selectivity of folded $Q$ bands and $CB^-$ bands [inset in Fig.~\ref{fig_2}(h)], the orbital angular momentum of $CB^-$ can be strongly altered and thus the valley Zeeman physics of the B exciton is more susceptible to changes. These unexpected proximity effects in the conduction bands of TMDs are also present in other van der Waals heterostructures. For instance, in TMDs coupled to ferromagnetic materials\cite{Zollner2019PRB_proximity, Zollner2020PRB, Ge2022npj, Zollner2022, Choi2022NatMat}, the proximity-induced exchange splitting is also quite complex and can be stronger in the conduction band depending on the particular geometry and stacking of the heterostructure.

We emphasize that the orbital angular momentum is not evaluated locally in the atomic spheres, but takes into account the whole spread of the wave function in the heterostructure, embedded in the transition matrix elements that enter the summation-over-bands expression of $L_z$ in Eq.~\ref{eq:Lz}. The orbital decomposition analysis in Fig.~\ref{fig_4} and Supplemental Note VI\cite{SM} provides a compact way of visualising the spreading of the wave function throughout the system and which bands from the original Gr and TMD layer are hybridizing in the heterostructure. In terms of perturbative approaches and effective models, the orbital decomposition analysis extracted from DFT provides valuable microscopic insight on the type of perturbation order and coupling mechanism behind the observed effects (such as changes in $L_z$ or $S_z$). For instance, a direct contribution of Gr states would encode some type of first order coupling, whereas the modification of the orbital decomposition within the TMD would encode higher order processes (virtual interlayer tunneling\cite{ David2019PRB, Peterfalvi2022PRR}). It is beyond the scope of this study to provide a full account of this physical phenomena in terms of effective models, such as performed in Refs.\cite{Li2019PRB, David2019PRB, Peterfalvi2022PRR}, but, instead, reveal the underlying microscopic picture within DFT, similarly to Refs.\cite{Naimer2021PRB, Pezo2021TDM}.

We have also analyzed the reduced exciton masses extracted directly from the DFT calculations, presented in Fig.~\ref{fig_5}(a), and found that the reduced mass for the A exciton barely changes while for the B exciton the reduced mass changes $~0.005 m_e$, which are still rather weak to be clearly visible in experiments\cite{Raja2019}. Moreover, we inspected the variations in the relevant energy scales $E_g$, $\Delta_c$ and $\Delta_v$ [defined in Fig.~\ref{fig_1}(b)]. We found that the band gap varies by $\sim \!\! 15$ meV, $\Delta_c$ varies $\sim \!\! 2$ meV and $\Delta_v$ varies $\sim \!\! 6$ meV for the different systems, which are relatively small changes. The associated correction to the $g$-factor due to variations in $E_g$ is on the order of $\Delta E_g / E_g \!\! \approx \!\! 10^{-2}$, certainly smaller than the observed changes in Fig.~\ref{fig_3}. These small contributions originating from the energy parameters provide further support to our picture that changes in $g_B$ are indeed manifestations of the complex nature of wave function hybridization across the different layers in the van der Waals heterostructure\cite{Slobodkin2020PRL, Plankl2021NatPhot, Barre2022Science}. Finally, we note that effective masses and transition matrix elements are essentially unaffected over the different WS$_2$/Gr systems investigated[see Figs.~\ref{fig_2}(i,j) and Fig.~\ref{fig_5}(a)], providing further support to the perspective that Gr alters the exciton binding energies on the dielectric level\cite{Stier.2016b, Raja2017, Waldecker2019PRL} (see also discussions in Section II and Supplemental Note III\cite{SM}).

\begin{figure}[htb]
\begin{center} 
\includegraphics{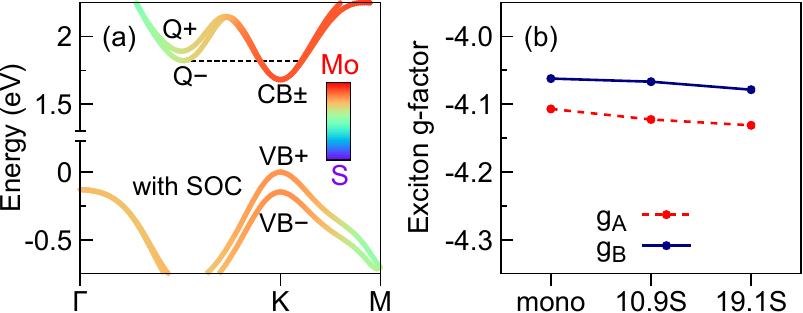}
\caption{(a) Band structure of MoS$_2$ monolayer showing the spread of the wave function over Mo and S atomic spheres with SOC. The dashed horizontal line starting at the Q$-$ band indicate the off-resonant energy condition to the $CB^\pm$ bands. (b) Calculated exciton g-factors, $g_A$ and $g_B$, for the different MoS$_2$/Gr systems considered.}
\label{fig_6}
\end{center}
\end{figure}

With this microscopic understanding, it would be interesting to consider a limiting scenario in which the modification of the valley Zeeman physics mediated by the $Q$ point folded bands is essentially suppressed, thus isolating the direct contribution of Gr. Unlike monolayer WS$_2$, shown in Fig.~\ref{fig_3}(a-c), in Mo-based TMD monolayers the $Q$ point conduction bands above the $K$ valleys\cite{Kormanyos2015TDM, Zollner2019PRB_strain} and therefore we expect this resonant coupling to be strongly suppressed in MoX$_2$/Gr (X=S,Se) systems. In order to verify this hypothesis and strengthen our understanding discussed above, we have analyzed the 10.9S and 19.1S cases for MoS$_2$/Gr heterostructures and summarize our findings in Fig.~\ref{fig_6} (with further details given in Supplemental Note VIII\cite{SM}). Because the $Q$ bands (either $Q^-$ or $Q^+$) are energetically out-of-resonance with respect to the $CB^\pm$ bands [Fig.~\ref{fig_6}(a)], we found no significant changes to the exciton $g$-factors when comparing the monolayer with the 10.9S and 19.1S MoS$_2$/Gr heterostructures, shown in Fig.~\ref{fig_6}(b), in line with our understanding that $Q$ point bands are crucial to mediate the interaction between Dirac cone of Gr and the $K$ point bands of the TMD.


\section*{IV. Modified valley Zeeman physics by graphene as a dielectric}

To complete our analysis, we investigate the modification of the valley Zeeman physics of WS$_2$ monolayer due to the influence of Gr on the dielectric level, motivated by the fact that the binding energy, $E_B$, and the exciton emission energy can be well explained in terms of dielectric screening effects\cite{Stier.2016b, Raja2017, Waldecker2019PRL} (see also Section II and Supplemental Note III\cite{SM}). There are two relevant effects related to the influence of the dielectric screening on the exciton emission energy: (i) the band gap reduction (rigid band shift)\cite{Cho2018PRB, Waldecker2019PRL}, which is $\sim \!\! 0.15 (0.3)$ eV\cite{Raja2017} in the Gr/WS$_2$ (Gr/WS$_2$/Gr) case; and (ii) the localization of the exciton wave function in $k$-space (increase of the exciton radius in real space), which can be directly extracted from our calculations (see Table III in Supplemental Note IV\cite{SM} for the changes in the FWHM of the exciton envelope function). We note that recent experiments have shown that the measured band and exciton $g$-factors agree remarkably well with density functional theory (DFT) calculations without excitonic corrections\cite{Robert2021PRL, Zinkiewicz2021NL}. However, it is in principle expected that excitonic effects renormalize the exciton $g$-factors by averaging the spin and orbital degrees of freedom around the $K$ valleys\cite{Deilmann2020PRL, Chen2019NL, Gillen2021, Heissenbuttel2021NL, Amit2022, Hotger2022}. In order to provide a cohesive and comprehensive picture of the valley Zeeman physics, we explore the consequences of the band gap reduction and wave function localization by considering a pristine monolayer WS$_2$ (with a lattice parameter of 3.153 $\textrm{\AA}$ and thickness of 3.14 $\textrm{\AA}$\cite{Kormanyos2015TDM}, without any relaxation from the heterostructure). The DFT calculations provide the fundamental information for the $g$-factors and effective masses of the low-energy bands (see Fig.~\ref{fig_1}(b) for the relevant bands). The exciton wave functions are calculated within the effective Bethe-Salpeter equation in $k$-space using the DFT effective masses as input (details in Supplemental Note IV\cite{SM}). The computational details for the DFT calculations can be found in the Supplemental Note V\cite{SM}.

The band gap reduction due to the increased dielectric screening can be taken into account by applying a scissor shift to the calculated DFT band gap, i. e., a negative rigid energy shift to all conduction bands. These shifts are then incorporated in the energy differences that appear in the expression for the orbital angular momentum in Eq.~\ref{eq:Lz}. In Table~\ref{tab:g_scissor}, we present the calculated exciton $g$-factors for scissor shifts of $0$, $-0.2$ and $-0.4$ eV. The $g$-factors $g_A$ and $g_B$ become $\sim \!\! 0.1$ more negative as we decrease the band gap by $-0.4$ eV. On the other hand, the difference $g_A - g_B$ is rather small ($\sim \!\! 10^{-2}$) and barely changes with the applied scissor shift. Therefore, the reduced band gap due to increased dielectric screening in WS$_2$/Gr heterostructures is unlikely to drastically modify the quantity $g_A - g_B$. We point out that our calculated $g$-factors given in Table~\ref{tab:g_scissor} are in excellent agreement with the experimental values given Table~\ref{tab:g_exp} for the WS$_2$ case, which is consistent with the previous reports comparing DFT-based calculations with experimentally determined band and exciton $g$-factors in WSe$_2$\cite{Robert2021PRL} and WS$_2$\cite{Zinkiewicz2021NL}.

\begin{table}[htb]
\caption{Calculated exciton $g$-factors for pristine WS$_2$ monolayer within DFT for different values of scissor shifts to simulate the effect of band gap reduction.}
\begin{center}
\begin{tabular}{cccc}
\hline
\hline
   & \multicolumn{3}{c}{Scissor shift (eV)}\tabularnewline
              &   $0.0$ &  $-0.2$  & $-0.4$\tabularnewline
\hline
$g_{A}$       & $-3.72$ & $-3.77$ & $-3.82$\tabularnewline
$g_{B}$       & $-3.74$ & $-3.79$ & $-3.85$\tabularnewline
$g_{A}-g_{B}$ &  $0.02$ &  $0.02$ &  $0.03$\tabularnewline
\hline
\hline
\end{tabular}
\end{center}
\label{tab:g_scissor}
\end{table}

\begin{table}[htb]
\caption{Calculated exciton $g$-factors for pristine WS$_2$ monolayer including the effects of the exciton wave function localization in $k$-space. The calculations with DFT effective masses and corrected (Corr.) masses ($\sim \!\! 13\%$ increase to reach experimental values) are shown.}
\begin{center}
\begin{tabular}{ccccc}
\hline
\hline
 &  & WS$_2$ & Gr/WS$_{2}$ & Gr/WS$_{2}$/Gr\tabularnewline
 \hline
DFT  &       $g_{A}$ & $-3.21$ & $-3.37$ & $-3.46$\tabularnewline
mass &       $g_{B}$ & $-3.15$ & $-3.34$ & $-3.44$\tabularnewline
     & $g_{A}-g_{B}$ & $-0.06$ & $-0.03$ & $-0.02$\tabularnewline
 \hline
Corr. &       $g_{A}$ & $-3.13$ & $-3.32$ & $-3.42$\tabularnewline
mass  &       $g_{B}$ & $-3.06$ & $-3.28$ & $-3.39$\tabularnewline
      & $g_{A}-g_{B}$ & $-0.07$ & $-0.04$ & $-0.03$\tabularnewline
 \hline
 \hline
\end{tabular}
\end{center}
\label{tab:g_exc}
\end{table}

The effect of the dielectric screening on the spatial extension of the exciton wave function and how it translates to the exciton $g$-factor can be investigated by the following relation\cite{Deilmann2020PRL, Heissenbuttel2021NL, Amit2022}
\begin{equation}
g_X = 2 \int \textrm{d}\vec{k} \left[g_z(c,\vec{k}) - g_z(v,\vec{k}) \right] \left| F_{c,v}(\vec{k}) \right|^2
\end{equation}
in which $X$ is the exciton state generated by the conduction band, $c$, and valence band, $v$, described by the envelope function $F_{c,v}(\vec{k})$, with wave vector $\vec{k}$ centered at the $K$ valley. Since the 1s exciton envelope function in our approximation has a radial symmetry, we also consider such dependence for the $g$-factors, as discussed in Supplemental Note IX\cite{SM}. We summarize our results in Table~\ref{tab:g_exc}. The absolute values of both $g_A$ and $g_B$ become less negative, providing a worse comparison with experiments (considering the WS$_2$ case as reference, shown in Table~\ref{tab:g_exp}), but consistent with the renormalization effects observed in previous theoretical works\cite{Chen2019NL,Deilmann2020PRL}. More importantly, the difference $g_A - g_B$ remains quite small and barely changes with the increased dielectric confinement. Furthermore, the calculated values of $g_A$ and $g_B$ are consistent whether we use the DFT masses or the corrected values (increased by $\sim \!\! 13 \%$ to approach experimental values\cite{Goryca.2019}). Interestingly, the calculated g-factor change from the WS$_2$ to the Gr/WS$_{2}$/Gr case is 0.25 (0.29) for the A (B) excitons, which is in line with the experimental value of 0.29 (0.33). Our analysis suggest that the observed changes in $g_A - g_B$ do not originate from the dielectric effect on the exciton wave function.

Besides the dielectric effects, it is also worth mentioning the possible role of strain that could be present in the WS$_2$/Gr heterostructures due to the growth process. Typically, the reminiscent strain from the fabrication procedure was be estimated to be on the order of $\pm0.5\%$\cite{Mennel2018NatComm, Darlington2020NatNano}. Furthermore, recent calculations\cite{FariaJunior2022NJP} have shown that in monolayer TMDs, $g_A$ and $g_B$ do not deviate from each other unless there is a sizable amount of compressive strain $>$ 2\%, leading to $g_A - g_B > 0$, which is accompanied with a reduction of the magnitude in both g-factors, i. e., $g_A$ and $g_B$ become less negative. This level of strain is very unlikely to be present in our samples, as this would be easily noticeable in the exciton energy (shifts due to strain are typically on the order of 100 meV/\%)\cite{Zollner2019PRB_strain, Frisenda2017npj2DMA}.

We wrap up our theoretical analysis by emphasizing that the calculated dielectric effects of band gap reduction and exciton wave function localization are able to explain the monotonic behavior of the observed $g$-factors (they become more negative as the dielectric confinement increases). On the other hand, the drastic changes observed in $g_A-g_B$ (from negative to positive) can only be explained by our first principles calculations (Section III) including the microscopic effect of the interlayer coupling between WS$_2$ and graphene, which ultimately translates to changes in $g_B$ originating from the orbital degree of freedom of the lower conduction band $CB^-$.


\section*{V. Concluding remarks}

In summary, we reveal signatures of proximity-enhanced valley Zeeman physics, detected through high field magneto optical spectroscopy on the A- and B-exciton of Gr-encapsulated ML WS$_2$. We investigate the markedly different evolution of the $X^A$ and $X^B$ $g$-factors through first principles calculations by exploring different prototypical scenarios of the twist angle, stacking and material composition. We reveal that the proximity effect due to the Gr-encapsulation mainly appears in the lowest $K$-point CB of WS$_2$ ($CB^-$) due to the modification of the orbital angular momentum, while the spin angular momentum is nearly unaffected. We reveal a mechanism, where the hybridization of the CB $Q$-point with $CB^-$ is mediated by the Gr Dirac cone and enhanced through commensurate stacking and the energetic alignment of the conduction bands at the $Q$- and $K$-point. 

We expect our results to be general for van der Waals and moiré heterostructures, also affecting spin dynamics at such interfaces. Particularly, we show how sensitive the $g$-factor is to the detailed nuances of the wave function hybridization, ultimately translated to the orbital degree of freedom in a non-intuitive manner. We therefore expand the concept of proximity effects in these heterostructures and show how magnetooptical experiments combined with first principles calculations can be used as a tool to quantify wave function hybridization. Future work may exploit the possibility to resonantly tune the coupling across layers via external electric fields, long-range moiré scales and mechanical deformations.


\ack
We thank Klaus Zollner and Marko Milivojevic for helpful discussions. Work at the National High Magnetic Field Laboratory (NHMFL) was supported by National Science Foundation (NSF) Grant No. DMR-1157490, the State of Florida, and the U.S. Department of Energy (DOE). P.E.F.J., T.N. and J.F. acknowledge the financial support of the Deutsche Forschungsgemeinschaft (DFG, German Research Foundation) SFB 1277 (Project-ID 314695032, projects B07 and B11) and of the European Union Horizon 2020 Research and Innovation Program under Contract No. 881603 (Graphene Flagship). P.E.F.J., T.N., J.F., J.J.F and A.V.S. acknowledge the financial support of SPP 2244 (Project No. 443416183). 


$\;$

\bibliography{biblio}


\end{document}


\title{Supplemental Material for ``Proximity-enhanced valley Zeeman splitting at the WS$_2$/graphene interface''}

\author{Paulo E. Faria~Junior}
\email{paulo-eduardo.faria-junior@ur.de}
\affiliation{Institute for Theoretical Physics, University of Regensburg, 93040 Regensburg, Germany}

\author{Thomas Naimer}
\affiliation{Institute for Theoretical Physics, University of Regensburg, 93040 Regensburg, Germany}

\author{Kathleen M. McCreary}
\affiliation{Materials Science and Technology
Division, Naval Research Laboratory, Washington, Washington DC 20375, USA}

\author{Berend T. Jonker}
\affiliation{Materials Science and Technology
Division, Naval Research Laboratory, Washington, Washington DC 20375, USA}

\author{Jonathan J. Finley}
\address{Walter Schottky Institut and Physik Department, Technische Universität München, Am Coulombwall 4, 85748 Garching, Germany}

\author{Scott A. Crooker}
\affiliation{National High Magnetic Field Laboratory, Los Alamos, New Mexico 87545, USA}

\author{Jaroslav Fabian}
\email{jaroslav.fabian@ur.de}
\affiliation{Institute for Theoretical Physics, University of Regensburg, 93040 Regensburg, Germany}

\author{Andreas V. Stier}
\email{andreas.stier@wsi.tum.de}
\affiliation{Walter Schottky Institut and Physik Department, Technische Universität München, Am Coulombwall 4, 85748 Garching, Germany}
\affiliation{National High Magnetic Field Laboratory, Los Alamos, New Mexico 87545, USA}

\maketitle

\tableofcontents


\clearpage

\newpage

\section{Experimental setup}

Broadband white light from a xenon lamp was coupled
to the samples using a 100 mm diameter multimode optical fibre.
The light was focused onto the sample at near-normal incidence
using a single aspheric lens, and the reflected light was refocused
by the lens into a 600 mm diameter collection fibre. A thin-film
circular polarizer mounted over the delivery or collection
fibre provided $\sigma^{\pm}$ circular polarization sensitivity. During the experiment, the handedness of the ciruclar polarizer was kept constant and the the magnetic field direction was switched $\pm B$ to access both circular polarization helicities.
The collected light was dispersed in a 300mm spectrometer
and detected with a charge-coupled device detector. Broadband reflection spectra were acquired continuously at a rate of $\sim 500$ Hz throughout the $\sim$70 ms long magnet pulse. Details about the experiment can be found in\cite{Stier.2016c}.

\clearpage

\newpage

\section{Dielectric screening in 2D materials stacks}

Exciton Rydberg states in 2D TMDs have been successfully modeled with the use of the popular "Rytova-Keldysh" (RK) potential\cite{Rytova1967,Keldysh1979}, which describes the electrostatic potential between an electron and a hole in a thin semiconductor slab, confined between bulk dielectrics. In the limit of an infinitely thin 2D semiconductor, the RK potential can be expressed analytically in reciprocal space as

\begin{equation}
V_{RK}(q)=\frac{2 \pi e^2}{q}\frac{1}{\kappa+r_0q}
\label{eq:VRK}
\end{equation}

where the dielectric constant of the bulk materials around the 2D semiconductor, $\kappa=\frac{\kappa_1+\kappa_2}{2}$, and the screening length of the 2D semiconductor  $r_0=2\pi\chi_{2D}$ is defined from the 2D optical susceptibility. As such both the material surrounding the 2D slab and the susceptibility of the slab describe the $q$-dependence of this potential. For WS$_2$, the 2D susceptibility is given by $r_0=2\pi\chi_{2D}=\frac{d(\epsilon_{\perp}-1)}{2}\approx\frac{d\epsilon_{\perp}}{2}$, where $d=c/2$ is half the lattice constant perpendicular to the 2D plane ($c_{WS_2}=12.35 \; \textrm{\AA}$) and the dielectric constant perpendicular to the plane is $\epsilon_{\perp}=13.3$. As such, for freestanding WS$_2$, $r_0=2\pi\cdot6.03 \; \textrm{\AA}=3.8 \; \textrm{nm}$\cite{Berkelbach2013}. 

Encapsulation of the 2D semiconductor with a bulk dielectric, as shown in Fig. SI1 (a), modifies both $r_0\rightarrow r_{0,bulk}=\frac{d(\epsilon_{\perp}-1)}{2}(1-\frac{\kappa_1^2+\kappa_2^2}{2\epsilon_{\perp}^2})$ and $V_{RK}(q)=\frac{2 \pi e^2}{q}\frac{1}{\frac{\kappa_1+\kappa_2}{2}+r_{0,bulk}q}$. This explains, for example, the recent experimental parametrization of $V_{RK}(q)$ and $r_{0,bulk}=3.4 \; \textrm{nm}$ for WS$_2$ encapsulated in bulk hBN ($\kappa_{hBN}=4.35$)\cite{Goryca.2019}. The encapsulation of a monolayer TMD with individual layers of graphene and successive increased dielectric screening and reduction of the single particle bandgap has been parameterised with an effective dielectric constant $\kappa$\cite{Raja2017} via the 1s-2s splitting of exciton Rydberg states ($\Delta_{12}$). The accuracy of this method, however, relies on a known relationship of $\Delta_{12}$ on the exciton binding energy, which for a given magnitude of the exciton binding energy can vary significantly for different parameter pairs of ($\kappa$,$r_0$). Specifically for the encapsulation with atomically thin materials, the Coulomb interactions are re-normalized through the screening length of the 2D slab and not through a re-normalization of the bulk dielectric $\kappa$. As such, for a ML WS$_2$  in a layer of graphene, as shown in Fig. SI1 (b), the screening length is re-normalized as $r_0\rightarrow r_{0,g}=\frac{d(\epsilon_{\perp}-1)}{2}(1-\frac{\kappa_1^2+\kappa_2^2}{2\epsilon_{\perp}^2})+\frac{d_g\kappa_g}{2}(1-\frac{\kappa_2^2}{\kappa_g^2})$ and $V_{RK}(q) \rightarrow V_{RK,g}(q)=\frac{2 \pi e^2}{q}\frac{1}{\frac{\kappa_1+\kappa_2}{2}+r_{0,g}q}$. For the encapsulation with two monolayers of graphene, $r_0 \rightarrow r_{0,gg}=\frac{d(\epsilon_{\perp}-1)}{2}(1-\frac{\kappa_1^2+\kappa_2^2}{2\epsilon_{\perp}^2})+d_g\kappa_g(1-\frac{\kappa_2^2}{\kappa_g^2})$ and the respective modification of the Coulomb potential. As such, and in contrast to the encapsulation with a bulk material, the screening length is significantly enlarged to $r_{0,g}=59 \; \textrm{\AA}$ and $r_{0,gg}=81 \; \textrm{\AA}$ ($d_g=0.5 \; \textrm{nm}$ and $\kappa_g$=9). 

\begin{figure}[h!]
\begin{center} 
\includegraphics[width=0.76\textwidth]{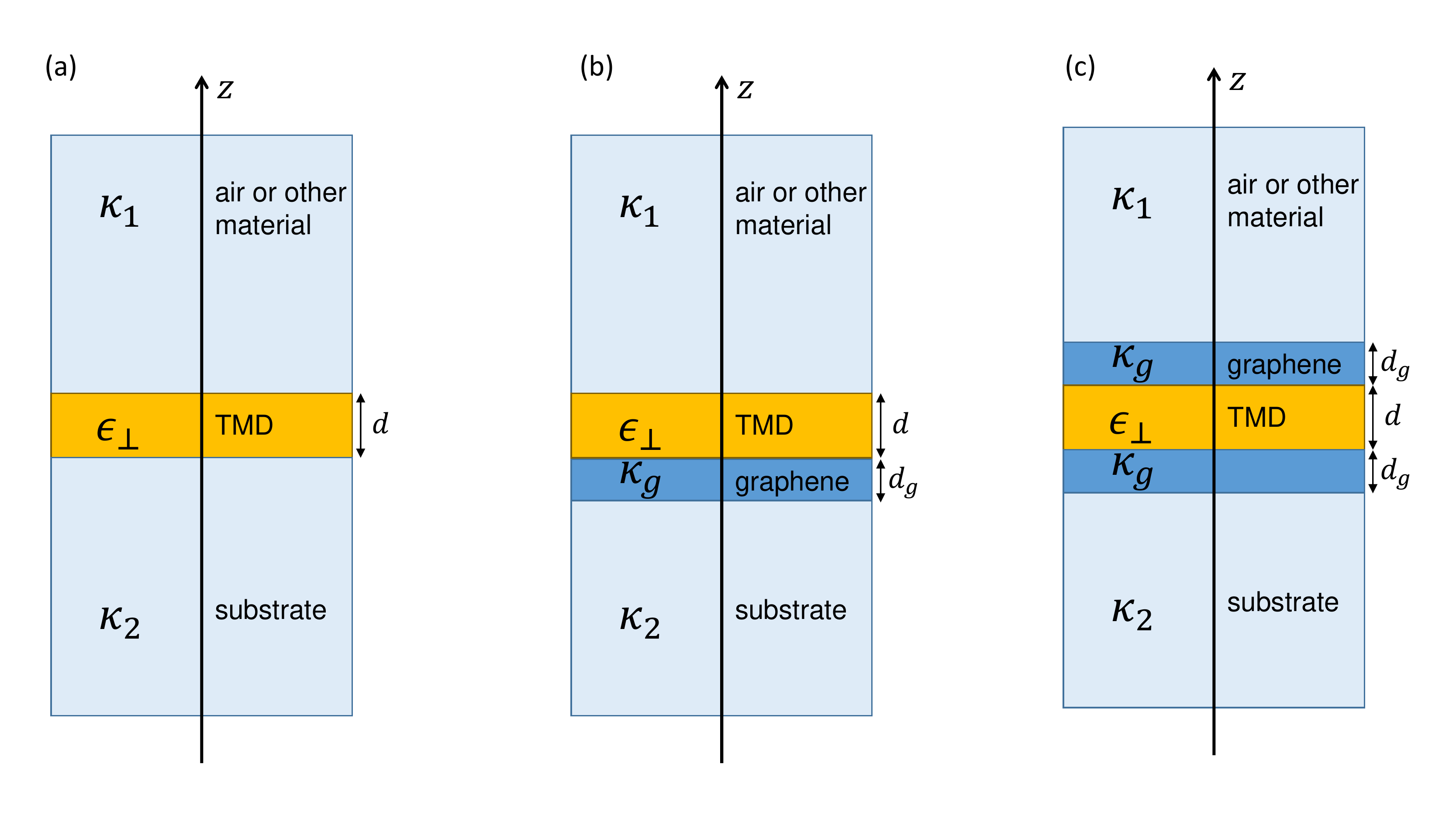}
\caption{Schematic of the dielectric environment of the three sample stacks investigated in this letter. (a) A monolayer TMD is placed on top of a bulk dielectric substrate (e.g. SiO$_2$ or fully encapsulated in hBN), (b) graphene below the TMD monolayer and (c) encapsulation with top and bottom monolayer graphene.}
\label{figSI1}
\end{center}
\end{figure}


\clearpage

\newpage

\section{Exciton binding energies from high field diamagnetic shift}

The consistency of this picture with our experimental high field data is depicted in Figure SI2\cite{Stier.2016, Stier.2018, Goryca.2019}. Knowledge of the diamagnetic shift $\sigma$ constrains the exciton radius $r_1$ when $m_r$ is known but also the exciton binding energy when the Coulomb potential $V_{RK}$ is known. For increased screening, the exciton binding energy is expected to decrease, therefore increasing the root mean squared radius of the 1s exciton $(r_1=\sqrt{\langle r^2 \rangle_{1s}})$. This is depicted in Figure SI2 (a), which shows that for increased encapsulation, the diamagnetic shift successively increases. Because excitons in 2D ML TMDs are small $(r_1\sim 1-2 ~nm)$, large magnetic fields are necessary to determine this small diamagnetic  shift $(\sigma \sim 0.1-0.3 ~\mu eV/T^2)$ reliably. The calculated and simulated exciton binding energies are shown in Figure SI2 (b,c), where we have utilized the Coulomb potential and parameters discussed in SI1 above.   
Figure SI2 (b) shows a color-coded surface plot of the exciton binding energy, calculated using the Fourier transformed $V_{RK}(r)$ over a range of reduced exciton masses and screening lengths. For the calculation, an average dielectric environment from the air/SiO$_2$ environment was used of $\kappa=\frac{1+2.25}{2}$.

\begin{figure}[h!]
\begin{center} 
\includegraphics[width=0.99\textwidth]{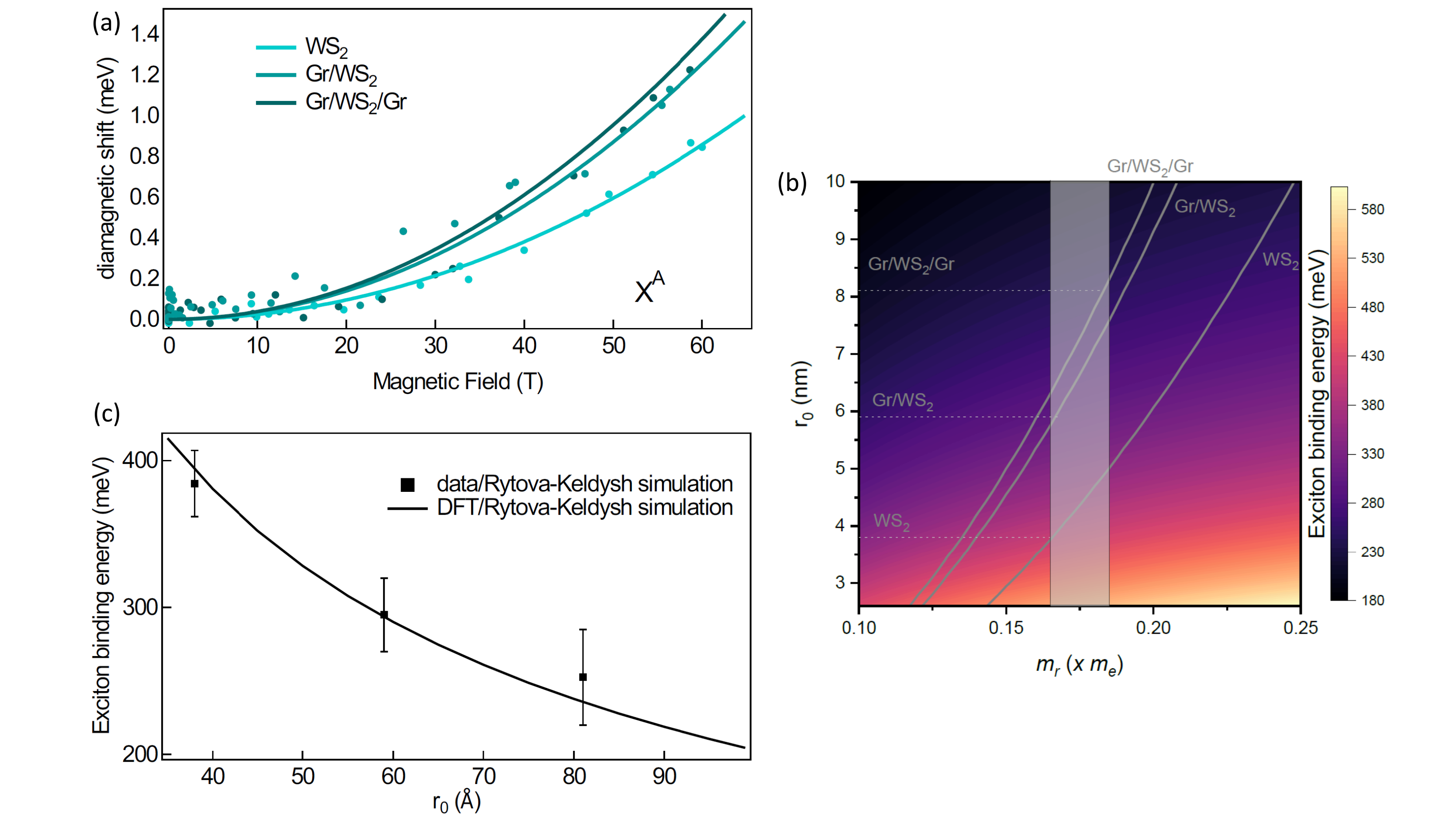}
\caption{(a) Average energy shift of the A-exciton for each sample stack. Solid lines are fits to $\sim B^2$, from which the diamagnetic shift coefficient $\sigma$ was determined. (b) Colour surface plot of the calculated exciton binding
energy, using $V_{RK}(q)$ from equation (1), over a range of reduced exciton mass $m_r$ and screening length $r_0$ for an effective dielectric screening of air and SiO$_2$ substrate of $\kappa=\frac{\kappa_1+\kappa_2}{2}=\frac{1+2}{2}$. Dashed lines indicate the screening length of the three sample stacks, described above.
Solid lines indicate contours of constant diamagnetic shift corresponding to
those measured in Fig. 1 of the main text for the A excitons in the respective sample stacks. The shaded box depicts the value for $m_r$ of monolayer WS$_2$ (including error bars). The exciton binding energy is deduced from the intersection of the diamagnetic shift with $r_0$ and $m_r$. (c) Deduced exciton binding energy from the diamagnetic shift data and calculated exciton binding energy calculated from a DFT/Bethe-Salpether approach.}
\label{figSI2}
\end{center}
\end{figure}

 At each point, we additionally computed the 1s exciton wavefunction $\Psi_{1s}(r)$, from which the expected diamagnetic shift coefficient $\sigma=e^2r_1^2/8m_r$ is determined. Indicated in the plot are solid contours of constant diamagnetic shift, as determined from the high field diamagnetic shift data, shown in Figure SI2 a. Because the reduced exciton mass for monolayer WS$_2$ was determined from investigation of the high field diamagnetic shift of $ns$-Rydberg states of hBN-encapsulated WS$_2$ (shaded area in Fig. SI 2.b)\cite{Goryca.2019}, the intersect of the calculated screening length for each sample stack and the diamagnetic shift contour marks the exciton binding energy, summarized in Table~\ref{tab:exc_exp}. The error bars of the deduced values are a combination of uncertainties in $m_r$ and $\sigma$. In Figure SI2 c, we show the deduced exciton binding energy using this method (scatter), together with the calculated exciton binding energy using a DFT/Bethe-Salpether approach, discussed below (solid line). 
 
\begin{table}[h!]
\caption{Summary of screening length and experimentally determined exciton binding energies.}
\begin{center}
\begin{tabular}{ccc}
 \hline
 \hline
 Material& Screening length ($\textrm{\AA}$) &Exciton binding energy (meV)\\
 \hline
 WS$_2$       &   37.5  &  384 $\pm$ 23 \\
 WS$_2$/Gr    &   59.4  &  295 $\pm$ 25 \\
 Gr/WS$_2$/Gr &   80.8  &  253 $\pm$ 33 \\
 \hline
\end{tabular}
\end{center}
\label{tab:exc_exp}
\end{table}


\clearpage

\newpage

\section{Exciton binding energies and wavefunctions}

The effective Bethe-Salpeter equation for excitons in k-space\cite{Rohlfing1998PRL, Rohlfing2000PRB, FariaJunior2019PRB, Zollner2019PRB_proximity,Zollner2019PRB_strain} assuming non-interacting parabolic bands for electrons and holes, reads
\begin{equation}
\left[\left(\frac{\hbar^{2}}{2m_{0}}\right)\frac{k^{2}}{\mu}-\Omega_{\lambda}\right]\,F_{\lambda}(\vec{k})-\sum_{\vec{k}^{\prime}}\text{V}_{\text{RK}}(\vec{k}-\vec{k}^{\prime})\,F_{\lambda}(\vec{k}^{\prime})=0\,,
\end{equation}
with $\Omega_{\lambda}$ and $F_{\lambda}(\vec{k})$ being the exciton binding energy and wavefunction for the $\lambda$-th state, respectively, and the exction reduced mass is written as $\mu^{-1}=m_{c}^{-1}+m_{v}^{-1}$, with $m_{c(v)}$ being the effective mass of the conduction (valence) band. The RK potential is given in Eq.~\ref{eq:VRK} with $r_0$ given in Table~\ref{tab:exc_exp} and $\kappa = 1.625$.

\begin{table}[h!]
\caption{Exciton masses obtained from the conduction and valence band masses calculated via DFT and with the correction of $\sim 13\%$ to reach the reported experimental values\cite{Goryca.2019}.}
\begin{center}
\begin{tabular}{ccc}
\hline
\hline
 & A exciton & B exciton\tabularnewline
\hline
DFT mass & 0.1543 & 0.2090\tabularnewline
Corr. mass & 0.1750 & 0.2374\tabularnewline
\hline
\hline
\end{tabular}
\end{center}
\label{tab:masses}
\end{table}

\begin{table}[h!]
\caption{Exciton binding energies and the full width at half maximum (FWHM) of the exciton envelope function squared, i. e., $\left|F_{\lambda}(\vec{k})\right|^2$.}
\begin{center}
\begin{tabular}{ccccccccc}
\hline
\hline
 &  & \multicolumn{3}{c}{A exciton} &  & \multicolumn{3}{c}{B exciton}\tabularnewline
\hline
 &  & WS$_2$ & Gr/WS$_2$ & Gr/WS$_2$/Gr &  & WS$_2$ & Gr/WS$_2$ & Gr/WS$_2$/Gr\tabularnewline
 \hline
\multirow{2}{*}{DFT mass} & $E_{b}(\textrm{meV})$ & 379.97 & 280.64 & 227.29 &  & 422.04 & 308.98 & 248.89\tabularnewline
 & FWHM $(\textrm{\AA}^{-1})$ & 0.0485 & 0.0405 & 0.0355 &  & 0.0580 & 0.0480 & 0.0420\tabularnewline
\hline
\multirow{2}{*}{Corr. mass} & $E_{b}(\textrm{meV})$ & 397.14 & 292.23 & 236.13 &  & 439.84 & 320.95 & 257.98\tabularnewline
 & FWHM $(\textrm{\AA}^{-1})$ & 0.0525 & 0.0435 & 0.0380 &  & 0.0625 & 0.0515 & 0.0450\tabularnewline
 \hline
 \hline
\end{tabular}
\end{center}
\label{tab:exc_calc}
\end{table}


\clearpage

\newpage

\section{First-principles calculations}

All heterostructures used for the calculations were initially created with the {\tt atomic simulation environment (ASE)}\cite{ASE} code using the lattice constants $a_{\textrm{Gr}}=2.46~\textrm{\AA}$ for graphene\cite{Gmitra2009PRB}, $a_{\textrm{WS}_2}=3.153~\textrm{\AA}$ for WS$_2$ with initial thickness of 3.14 $\textrm{\AA}$\cite{Kormanyos2015TDM} and $a_{\textrm{MoS}_2}=3.16~\textrm{\AA}$ for MoS$_2$ with initial thickness of 3.17$\textrm{\AA}$\cite{Kormanyos2015TDM}. In order to generate commensurate heterostructure supercells it is necessary to strain either one of the two components. Since we are more interested in the properties of TMDC, all strain is transferred to graphene, while the TMDC monolayer remains unstrained. This means the unstrained graphene lattice constant $a_{\textrm{Gr}}$ is only used in so far as to calculate the imposed strain. The atomic coordinates within the unit cell were then adjusted in a structural relaxation using the density functional theory (DFT) within the framework of {\tt Quantum ESPRESSO}\cite{Giannozzi2009:JPCM}. The self-consistent calculation for the structural relaxations converged when the forces were reduced below $10^{-5} \text{Ry}/a_0$, where $a_0$ is the Bohr radius, and a 9$\times$9 k-grid was used.  We used charge density and kinetic energy cutoffs ($E_\rho=550~\text{Ry}$ and $E_{wfc}=60~\text{Ry}$) for the scalar relativistic pseudopotentials with the projector augmented wave method\cite{Kresse1999:PRB} with the Perdew-Burke-Ernzerhof exchange-correlation functional\cite{Perdew1996PRL} and van der Waals effects corrections\cite{Grimme2006:JCC, Grimme2010:JCP, Barone2009:JCC}. A summary of the important structural parameters is shown in Table~\ref{tab:struct_WS2Gr}.

\begin{table}[h]
\caption{Structure information of the calculated WS$_2$/Gr and MoS$_2$/Gr heterostructures. For the definition of the integer pair (n,m) see Ref.~\cite{Naimer2021PRB}.}
\label{tab:struct_WS2Gr}
\begin{center}
\begin{tabular}{ccccccccc|cc}
\hline
\hline
 & \multicolumn{7}{c}{WS$_{2}$} &  & \multicolumn{2}{c}{MoS$_{2}$}\tabularnewline
 & 10.9S & 10.9E & 19.1S & 19.1S' & 19.1E & S2AB & S2AB &  & 10.9S & 19.1S\tabularnewline
 \hline
Angle & 10.893 & 10.893 & 19.107 & 19.107 & 19.107 & 19.107 & 19.107 &  & 10.893 & 19.107\tabularnewline
$\epsilon_{\text{Gr}}(\%)$ & $-$2.108 & $-$2.108 & $-$3.112 & $-$3.112 & $-$3.112 & $-$3.112 & $-$3.112 &  & $-$1.891 & $-$2.897\tabularnewline
(n,m) Gr & (2,1) & (2,1) & (2,1) & (2,1) & (2,1) & (2,1) & (2,1) &  & (2,1) & (2,1)\tabularnewline
(n,m) TMDC & (2,1) & (2,1) & (2,0) & (2,0) & (2,0) & (2,0) & (2,0) &  & (2,1) & (2,0)\tabularnewline
d-TMDC $(\textrm{Å})$ & 3.150 & 3.147 & 3.149 & 3.150 & 3.147 & 3.149 & 3.149 &  & 3.138 & 3.137\tabularnewline
TMDC-Gr $(\textrm{Å})$ & 3.290 & 3.287 & 3.290 & 3.293 & 3.286 & 3.273 & 3.273 &  & 3.357 & 3.359\tabularnewline
Gr rippling $(\textrm{Å})$ & 0.104 & 0.105 & 0.022 & 0.035 & 0.022 & 0.021 & 0.022 &  & 0.079 & 0.017\tabularnewline
$L_{\text{sc}}\;(\textrm{Å})$ & 8.342 & 8.342 & 6.306 & 6.306 & 6.306 & 6.306 & 6.306 &  & 8.361 & 6.320\tabularnewline
$N_{\text{atoms}}$ & 45 & 69 & 26 & 26 & 40 & 40 & 40 &  & 45 & 26\tabularnewline
\hline
\hline
\end{tabular}
\end{center}
\end{table}

The electronic structure, spin and g-factor calculations are performed using the WIEN2k package\cite{wien2k}, which implements DFT using the full potential all-electron scheme employing the augmented plane wave plus local orbitals (APW+lo) method. We used the Perdew-Burke-Ernzerhof (PBE) exchange-correlation functional\cite{Perdew1996PRL} with van der Waals corrections\cite{Grimme2010:JCP} for the heterostructures case, a Monkhorst-Pack k-grid of 6$\times$6 (12$\times$12) for the heterostructure (monolayer) and self-consistent convergence criteria of 10$^{-6}$ e for the charge and 10$^{-6}$ Ry for the energy. We considered a core–valence separation energy of $-6$ Ry, atomic spheres with orbital quantum numbers up to 10 and the plane-wave cutoff multiplied by the smallest atomic radii is set to 6.5. For the inclusion of SOC, core electrons are considered fully relativistically whereas valence electrons are treated in a second variational step\cite{Singh2006}, with the scalar-relativistic wave functions calculated in an energy window of -10 to 8 Ry. The upper energy window is particularly important to ensure the convergence of the orbital angular momentum, as shown in Refs.\cite{Wozniak2020PRB,FariaJunior2022NJP}.

\begin{figure}[h!]
\begin{center} 
\includegraphics{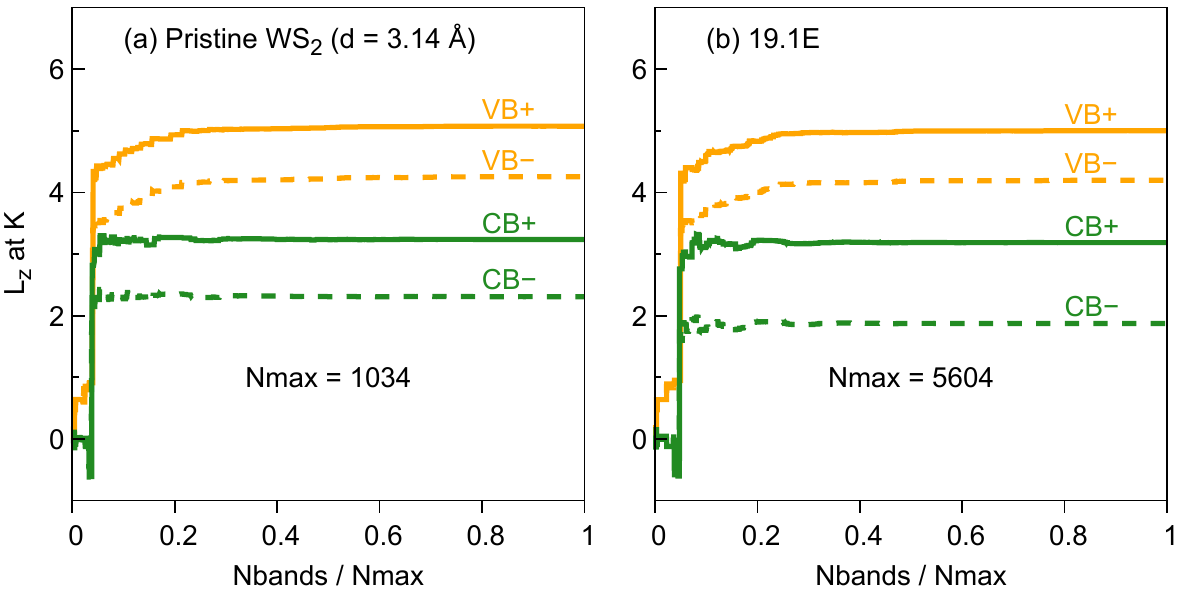}
\caption{Convergence of $\textrm{L}_{\textrm{z}}$ at the K valley as function of the number of bands for (a) the pristine monolayer WS$_2$ and (b) the Gr/WS$_2$/Gr heterostructure case 19.1E.}
\label{figSI3}
\end{center}
\end{figure}

In Fig.~\ref{figSI3} we show the convergence of $\textrm{L}_{\textrm{z}}$ at the K valley as function of the number of bands for pristine monolayer WS$_2$ and the Gr/WS$_2$/Gr heterostructure case 19.1E. There are many more bands in the heterostructure case in the same energy window ($\sim$ 5 times more than in the pristine monolayer) but the values of $\textrm{L}_{\textrm{z}}$ nicely converge.

In Table~\ref{tab:g_d} we compare the $g$-factors for different values of TMD thickness. There is a small change in the magnitude of the $g$-factors but $g_A - g_B$ remains nearly unchanged.

\begin{table}[h!]
\caption{Effect of thickness ($d$) on the $g$-factos of monolayer WS$_2$.}
\label{tab:g_d}
\begin{center}
\begin{tabular}{ccc}
\hline
\hline
 & $d = 3.140$ & $d = 3.151$ \tabularnewline
\hline
$g_{A}$ & $-3.72$ & $-3.67$ \tabularnewline
$g_{B}$ & $-3.74$ & $-3.68$ \tabularnewline
$g_{A}-g_{B}$ & $0.02$ & $0.01$ \tabularnewline
\hline
\hline
\end{tabular}
\end{center}
\end{table}


\clearpage

\newpage

\section{Orbital decomposition analysis}

\begin{figure}[h!]
\begin{center} 
\includegraphics{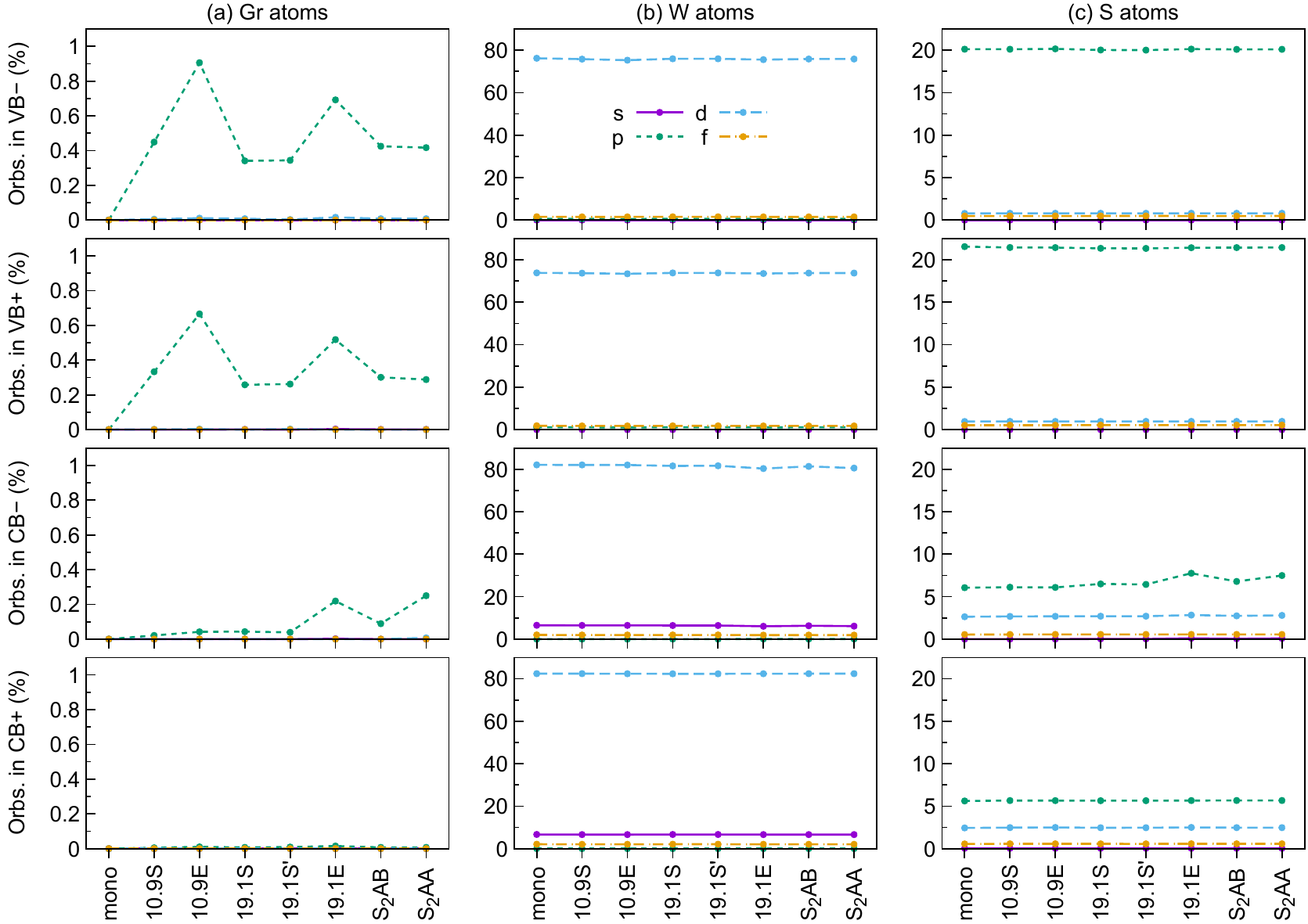}
\caption{Orbital decomposition of s, p, d and f-like characters within the atomic spheres of (a) Gr atoms, (b) W atoms and (c) S atoms.}
\label{figSI4}
\end{center}
\end{figure}

\begin{figure}[h!]
\begin{center} 
\includegraphics{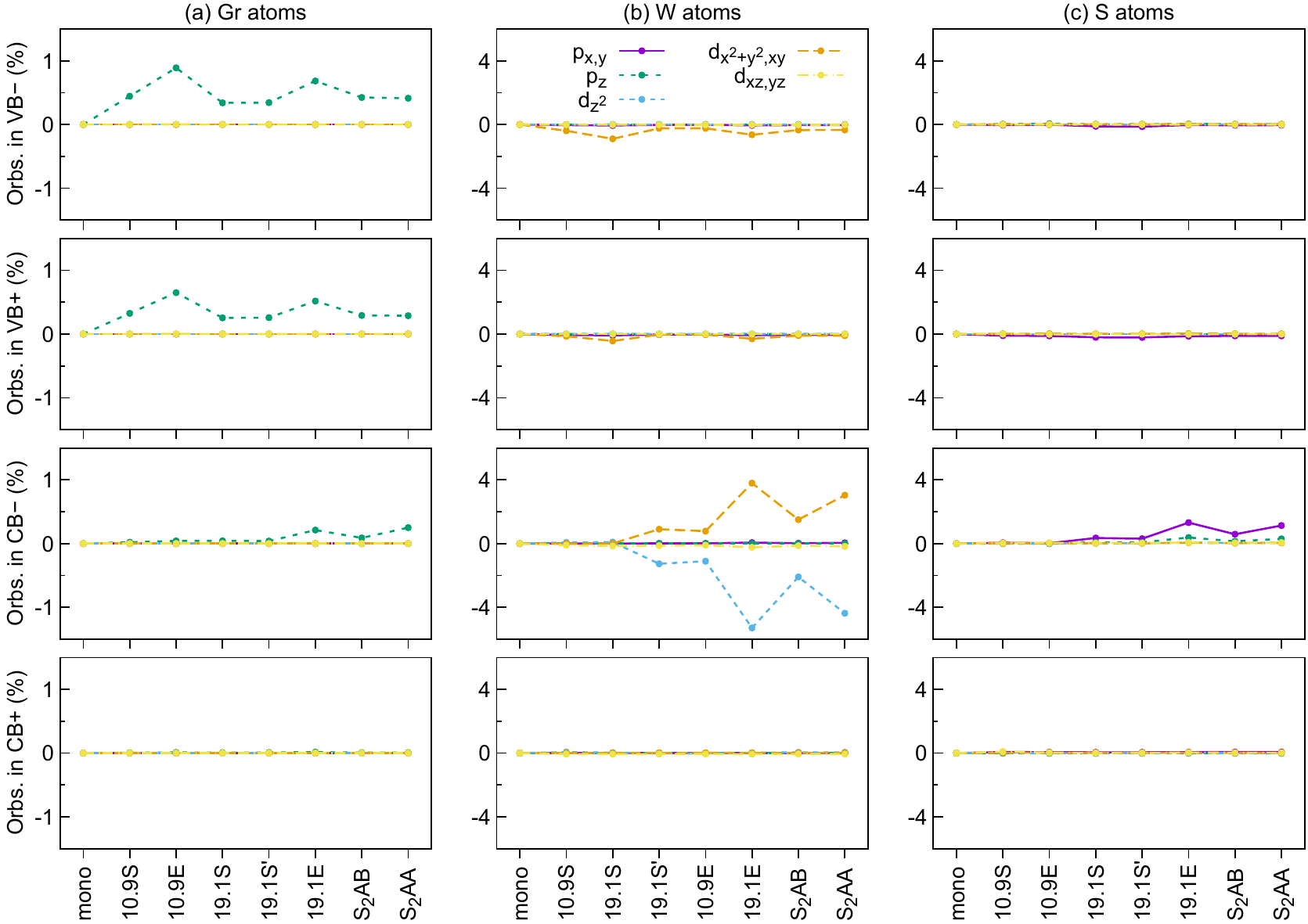}
\caption{Same as Fig.~\ref{figSI4} but showing only the contributions of p$_\textrm{x,y}$, p$_\textrm{z}$, d$_{\textrm{z}^2}$, d$_{\textrm{x}^2-\textrm{y}^2\textrm{,xy}}$, d$_\textrm{xz,yz}$ with respect to the value in the pristine monolayer. The decrease of d$_{\textrm{z}^2}$ character in CB$-$ becomes evident in this picture.}
\label{figSI5}
\end{center}
\end{figure}

In Fig.\ref{figSI4} and Fig.\ref{figSI5} we show the orbital decomposition for the van der Waals heterostructures. The decrease of dz2 character is accompanied with an enhancement of the d$_{\textrm{x}^2-\textrm{y}^2\textrm{,xy}}$ and p$_\textrm{x,y}$ characters, consistent with the contribution of $Q$-point bands.

\clearpage

\newpage

\section{Effect of lattice relaxation}

\begin{figure}[h!]
\begin{center} 
\includegraphics{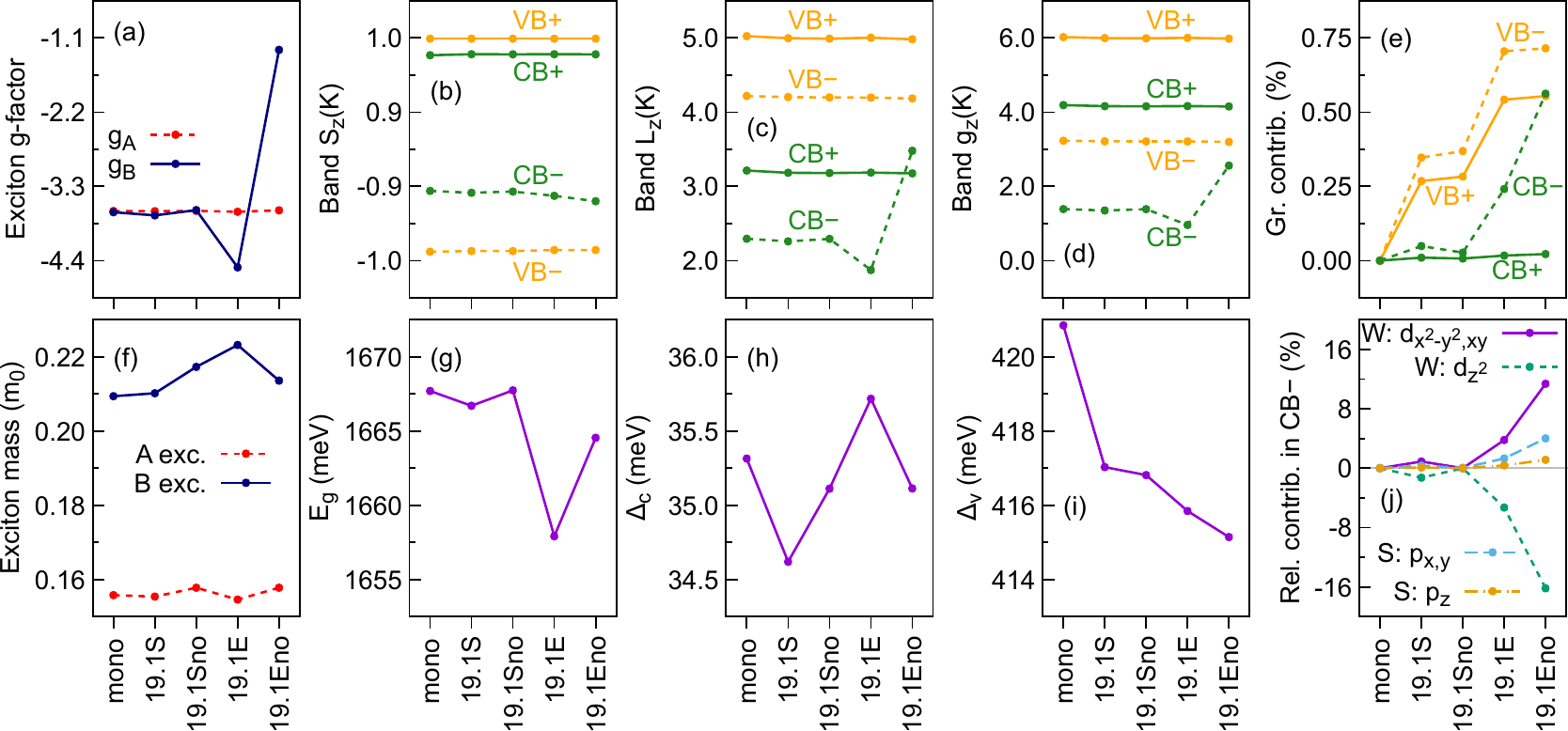}
\caption{(a) Calculated exciton g-factors, $g_A$ and $g_B$. (b) The spin expectation value $S_z$, (c) the orbital angular momenta, $L_z$, and (d) the g-factor, $g_z = L_z + S_z$, for CB$\pm$ and VB$\pm$. (e) Contribution of Gr atomic spheres to CB$\pm$ and VB$\pm$. (f) The effective masses for A and B excitons derived from first principles. The energy separations (g) $E_g$, (h) $\Delta_c$ and (i) $\Delta_v$, defined schematically in Fig.~1(b) of the main text. (j) Contributions of p$_\textrm{x,y}$, p$_\textrm{z}$, d$_{\textrm{z}^2}$, d$_{\textrm{x}^2-\textrm{y}^2\textrm{,xy}}$, d$_\textrm{xz,yz}$ with respect to the value in the pristine monolayer for the CB$-$ band.}
\label{figSI6}
\end{center}
\end{figure}

When dealing with van der Waals heterostructures, it is common to perform a structure relaxation in the system, which rearranges the atomic positions\cite{Gmitra2015PRB,Gmitra2016PRB,Pezo2021TDM, Yang2022Photonics} and generates rippling effects\cite{Naimer2021PRB}. We want to make sure that such rippling effects are not the main ingredient behind the observed changes to $g_B$. Essentially, we want to clarify that proximity-induced modification to $g$-factors would also happen in completely flat monolayers, ultimately stemming from the interlayer coupling. For this purpose, we calculated the 19.1S and 19.1E cases without atomic relaxation (using a WS$_2$ thickness of 3.151 $\textrm{\AA}$ and an interlayer distance of 3.274 $\textrm{\AA}$) and present the comparison with the relaxed case in Fig.~\ref{figSI6}. The same microscopic mechanism, discussed in Fig.~4 of the main text, applies here. Therefore, we confirmed that only $g_B$ is modified via the orbital degree of freedom of CB$-$ band and mediated by the $Q$-point folded bands. The absence of rippling effects provides a different interaction between WS$_2$ and Gr layers, thus leading to different values of $g_B$.

\clearpage

\newpage

\section{M\lowercase{o}S$_2$/G\lowercase{r} systems}

\begin{figure}[h!]
\begin{center} 
\includegraphics{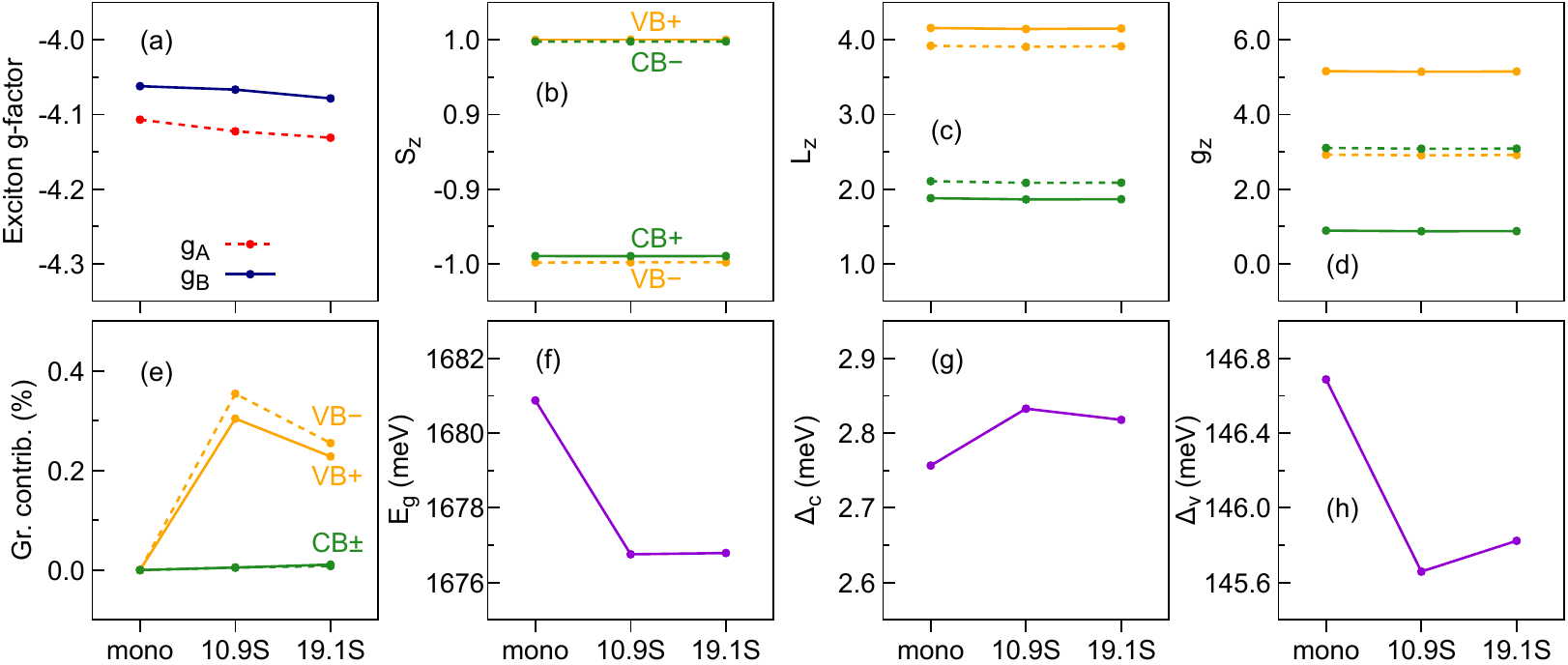}
\caption{(a) Calculated exciton g-factors, $g_A$ and $g_B$. (b) The spin expectation value $S_z$, (c) the orbital angular momenta, $L_z$, and (d) the g-factor, $g_z = L_z + S_z$, for CB$\pm$ and VB$\pm$. (e) Contribution of Gr atomic spheres to CB$\pm$ and VB$\pm$. The energy separations (f) $E_g$, (g) $\Delta_c$ and (h) $\Delta_v$, defined schematically in Fig.~1(b) of the main text.}
\label{figSI7}
\end{center}
\end{figure}

The calculations for MoS$_2$/Gr heterostructures are given in Fig.~\ref{figSI7}.

\clearpage

\newpage

\section{Excitonic effects on \lowercase{g}-factors}

\begin{figure}[h!]
\begin{center} 
\includegraphics{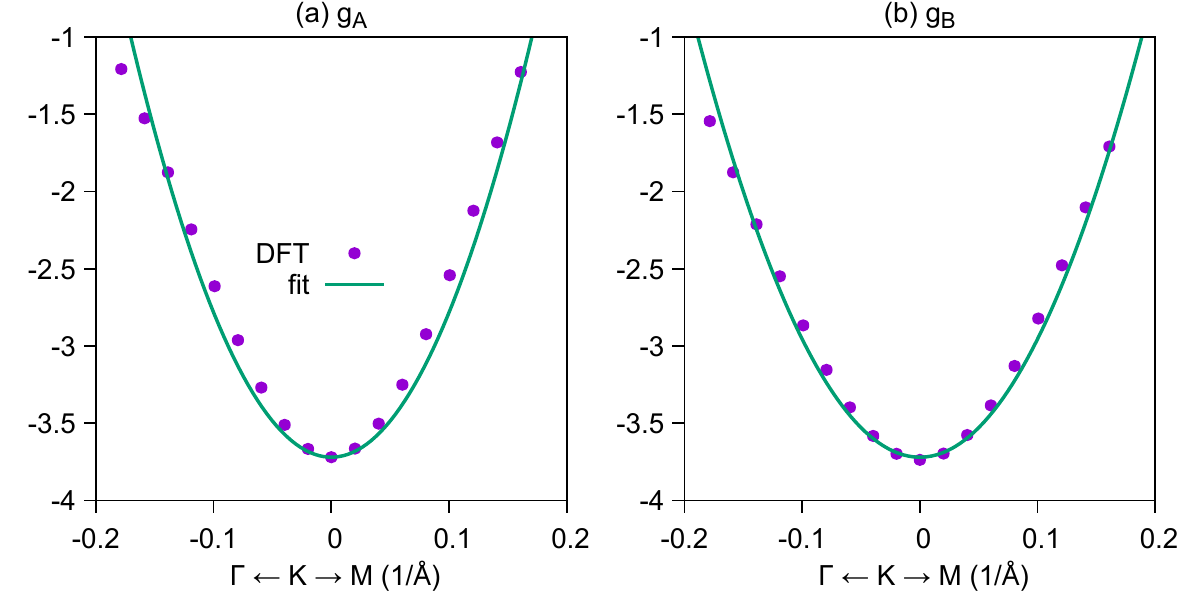}
\caption{DFT calculations and parabolic fitting $g_{X}(k)=g_{X}^{(0)}+g_{X}^{(2)}k^{2}$ for the (a) $g_A$ and (b) $g_B$ of pristine WS$_2$ monolayer. The quadratic fitting yield $g_{A}^{(2)} = 93.6308 \; \textrm{\AA}^{-2}$ and $g_{B}^{(2)} = 76.4669 \; \textrm{\AA}^{-2}$, with $g_{A}^{(0)} = -3.7197$ and $g_{B}^{(0)} = -3.7369$ (also given in the main text).}
\label{figSI8}
\end{center}
\end{figure}

To investigate the effects of the exciton wavefunction spreading in k-space in pristine WS$_2$ monolayer ($a_{\textrm{WS}_2}=3.153~\textrm{\AA}$ and $d_{\textrm{WS}_2}=3.14~\textrm{\AA}$), we calculated the spin and orbital angular momenta around the K valleys and obtained the $k$-dependence of the A and B exciton g-factors, shown in Fig.~\ref{figSI8} by the circles. In the region where the exciton wavefunction contributes the most, the $g$-factor has a convex dispersion with minimum at the K valley (consistent with previous calculations\cite{Deilmann2020PRL, FariaJunior2022NJP, Gobato2022NL, Kipczak2022}) and can be approximated by a parabola. The fitted parabola is shown in solid lines in Fig.~\ref{figSI8}. The influence of the exciton wavefunction on the g-factor renormalization is then computed via the expression
\begin{align}
g_{X} & =2\pi\int\text{d}k\,k\,g_{X}(k)\left|F(k)\right|^{2}\nonumber \\
 & =g_{X}^{(0)}+2\pi g_{X}^{(2)}\int\text{d}k\,k^{3}\left|F(k)\right|^{2} \, .
\end{align}

\clearpage

\newpage

$\;$

\bibliography{biblio}
